\documentclass[a4paper,11pt,notitlepage]{article}

\usepackage{geometry}
\usepackage{amsfonts,amsmath,amsthm,amssymb}
\usepackage{mathrsfs}
\usepackage[mathscr]{eucal}
\usepackage{color}
\usepackage{hyperref}
\usepackage{tikz}
\usetikzlibrary{shapes.geometric}
\usetikzlibrary{decorations.markings}
\usepgflibrary{decorations.shapes}
\usepgflibrary{shapes.misc}
\usepackage{authblk}

\usepackage{ctable}
\usepackage{array}

\newcommand{\dd}{\mathrm{d}}

\newcommand{\diag}{\mathrm{diag}}

\newcommand{\Hbulk}{\mathcal{H}_{\text{bulk}}}
\newcommand{\Hcontact}{\mathcal{H}_{\text{contact}}}
\newcommand{\Hout}{\mathcal{H}_{\text{out}}}
\newcommand{\Hin}{\mathcal{H}_{\text{in}}}
\newcommand{\UU}{U}
\newcommand{\me}{\mathrm{e}}
\newcommand{\TT}{\mathsf{T}}
\newcommand{\UUr}{U_{\text{r}}}
\newcommand{\UUs}{U_{\text{s}}}
\newcommand{\GL}{\mbox{GL}}

\newcommand{\id}{\mathbf{1}}
\newcommand{\gl}{\mathfrak{gl}}

\newcommand{\fk}{\mathfrak{k}}
\newcommand{\su}{\mathfrak{su}}

\newcommand{\pivac}{\pi_\text{vac}}
\newcommand{\psiin}{\psi_{\text{in}}}
\newcommand{\psiout}{\psi_{\text{out}}}

\DeclareMathOperator{\Tr}{Tr}
\DeclareMathOperator{\STr}{STr}
\DeclareMathOperator{\SDet}{SDet}
\DeclareMathOperator{\Det}{Det}

\title{Gaussian free fields at the \\ integer quantum Hall
  plateau transition}

\author[1]{R.~Bondesan\thanks{roberto.bondesan@phys.ox.ac.uk}}
\author[2]{D.~Wieczorek}
\author[2]{M.~R.~Zirnbauer}
\affil[1]{Rudolf Peierls Centre for Theoretical Physics, 1 Keble Road,
  Oxford OX1 3NP, UK}
\affil[2]{Institut f\"ur Theoretische Physik,
  Universit\"at zu K\"oln, Z\"ulpicher Stra{\ss}e 77, 50937 K\"oln,
  Germany}

\begin{document}
\date{December 13, 2016}

\maketitle

\begin{abstract}
In this work we put forward an effective Gaussian free field description of critical wavefunctions at the transition between plateaus of the integer quantum Hall effect. To this end, we expound our earlier proposal that powers of critical wave intensities prepared via point contacts behave as pure scaling fields obeying an Abelian operator product expansion. Our arguments employ the framework of conformal field theory and, in particular, lead to a multifractality spectrum which is parabolic. We also derive a number of old and new identities that hold exactly at the lattice level and hinge on the correspondence between the Chalker-Coddington network model and a supersymmetric vertex model.

\end{abstract}

\newpage

\tableofcontents

\newpage


\section{Introduction}

Among the critical phenomena in two-dimensional quantum systems with disorder, the transition between Hall conductance plateaus of the integer quantum Hall effect (referred to as the IQHE transition for short) stands out as a possible paradigm for quantum-phase transitions of Anderson-localization type. Yet, in spite of numerous efforts \cite{Bhaseen2000,Ikhlef2011,Bettelheim2012} and a renewed interest coming from the expanding research area of symmetry-protected topological phases \cite{Hasan2010}, the IQHE transition has so far defied an analytical solution by the methods of conformal field theory and/or the theory of integrable systems.

Due to wave interference by multiple scattering with random phases, a
striking feature of critical stationary states at the IQHE transition
is their multifractal nature. In the early going, analytical research
on the subject had assumed a parabolic law for the spectrum of
multifractal dimensions {\cite{Janssen1999, Bhaseen2000, Evers2001}}, in line with what happens for another, exactly solvable two-dimensional Anderson transition \cite{Kogan1996,MUDRY1996}. Later, several numerical
studies \cite{Evers2008b,Obuse2008} reported significant deviations
from parabolicity. If correct, these studies would rule out any
standard conformal field theory description of the IQHE transition.

The conundrum took a new perspective by a recent preprint by Suslov \cite{Suslov2014}. Based on a natural assumption, namely that the wave intensities $\vert \psi(r) \vert^2$ obey an operator product expansion of Abelian form, he suggested that the parabolic law for the multifractality spectrum should hold at Anderson transitions in general. While perhaps too far reaching to be true unconditionally, this suggestion has good merit in two dimensions, where the principles of conformal field theory constrain the possible scenarios. In the present paper, we put Suslov's suggestion on firm ground for the IQHE transition. The ultimate outcome will be that the (logarithm of the) field of wave intensities is distributed like a non-compact Gaussian free field with a background charge.

While the core of the argument is relatively short, we decided to seize this opportunity to write an expository paper, presenting a certain amount of material known to a small circle of experts but not the community at large. In doing so, we follow up on our recent proposal \cite{BWZ} to employ point contacts and so-called ``scattering states'' as the optimal tool for the study of the multifractality spectrum. Our motivation is that the wave intensities of these states have several attractive properties: they become independent of the system size in the large-volume limit, they show pure scaling behavior, and they do not suffer from termination of the multifractality spectrum.

The basis for our analytical arguments is the Chalker-Coddington network model \cite{Chalker1988} and its reformulation, by a variant of the Wegner-Efetov supersymmetry method, as a supersymmetric vertex model \cite{Zirnbauer1997}. Pioneered by N.\ Read, this variant introduces a Fock space for bosons and fermions to take disorder averages of products of retarded and advanced Green functions. We give an introduction to second quantization of the network model and derive some key formulas to use in this context. We also expound our ideas \cite{Gruzberg2013,BWZ} on how to construct pure-scaling fields from highest-weight vectors of the vertex model. Here an important technical point is that the Lie algebra of the non-compact sector has more than one conjugacy class of Cartan subalgebras, and one must choose the correct class to obtain highest-weight vectors that are positive and give rise to a continuum of scaling fields. To implement these highest-weight ideas, we derive new identities for a generating function relating wavefunction observables of the network model to correlation functions of the vertex model.

The paper is organized as follows. In Sect.\ \ref{sect:2} we describe our setup for the network model with point contacts. We define the notion of scattering states originating from point contacts and formulate the wavefunction observables of interest and their expressions in terms of Green functions. The procedure of second quantization of the network model, including a useful but rather unknown Gaussian integral representation due to Howe, is reviewed in Sect. \ref{ref:2nd}. In Sect.\ \ref{sect:4} we take the disorder average, passing from the network model to the vertex model. A long subsection, \ref{ref:nilpotent}, discusses the issue of Cartan subalgebras and positive highest-weight vectors. The relation between wavefunction observables of the network model and correlation functions of the vertex model is developed in Sect.\ \ref{sect:5}. Finally, Sect.\ \ref{sec:6} presents our arguments in favor of the effective description by a Gaussian free field. The paper ends with a Conclusion in \ref{sect:7}.


\section{The network model}\label{sect:2}

\subsection{Setup}

\begin{figure}
  \centering
  \begin{tikzpicture}[thick,decoration={ markings, mark=at position
    0.5 with {\arrow{>}}}, scale=1]
    \foreach \x in {2,4} 
    { 
      \foreach \y in {0,2} 
      {
        \draw[postaction={decorate}] (\x,\y)--(\x+1,\y+1);
        \draw[postaction={decorate}] (\x+1,\y+1)--(\x,\y+2); 
      } 
    }
    \foreach \x in {4} 
    { 
      \foreach \y in {0} 
      {
        \draw[postaction={decorate}] (\x,\y)--(\x-0.5,\y-0.5);
        \draw[postaction={decorate}] (\x+0.5,\y-0.5)--(\x,\y); 
      } 
      \foreach \y in {4} 
      {
        \draw[postaction={decorate}] (\x,\y)--(\x+0.5,\y+0.5);
        \draw[postaction={decorate}] (\x-0.5,\y+0.5)--(\x,\y); 
      } 
    }
    \foreach \x in {2} 
    { 
      \foreach \y in {2} 
      {
        \draw[postaction={decorate}] (\x-.5,\y+0.5)--(\x,\y);
        \draw[postaction={decorate}] (\x,\y)--(\x-0.5,\y-0.5); 
      } 
    }
    \foreach \x in {6} 
    { 
      \foreach \y in {2} 
      {
        \draw[postaction={decorate}] (\x,\y)--(\x+0.5,\y+0.5);
        \draw[postaction={decorate}] (\x+.5,\y-.5)--(\x,\y); 
      } 
    }
    \foreach \x in {3,5} 
    { 
      \foreach \y in {0,2} 
      {
        \draw[postaction={decorate}] (\x+1,\y+2)--(\x,\y+1);
        \draw[postaction={decorate}] (\x,\y+1)--(\x+1,\y); 
      } 
    }
    \begin{scope}[xshift=3cm,yshift=1cm]
      \draw[thick,red,<-] (0,0.5)[rounded corners=0.2cm] --(0.5,1) --
      (0,2-0.5);
      \draw[thick,red,->] (0.5,2)[rounded corners=0.2cm] --(1,1.5) --
      (1.5,2); 
      \draw[thick,red,->] (2,0.5)[rounded corners=0.2cm] --(1.5,1) -- (2,1.5);
      \draw[thick,red,<-] (0.5,0)[rounded
      corners=0.2cm] --(1,0.5) -- (1.5,0); 
      \draw[blue,fill=blue] (0.25,0.25) circle (0.1cm);
      \draw[blue,fill=blue] (2-0.25,2-0.25) circle (0.1cm);
    \end{scope}

    \begin{scope}[xshift=9cm]
      \begin{scope}[yshift=2.5cm]
        \draw[postaction={decorate}] (0,0)--(+1,+1);
        \draw[postaction={decorate}] (+1,+1)--(0,+2); 
        \draw[postaction={decorate}] (2,2)--(1,1);
        \draw[postaction={decorate}] (1,1)--(2,0); 
        \node at (-.5,1) { $A$:};
        \node at (0.5,2-.25) {{\small $3$}};
        \node at (0.5+1,2-.25) {{\small $2$}};
        \node at (1.5,.25) {{\small $4$}};
        \node at (1.5-1,.25) {{\small $1$}};
      \end{scope}
      \begin{scope}[yshift=-.5cm]
        \draw[postaction={decorate}] (0,2)--(+1,+1);
        \draw[postaction={decorate}] (+1,+1)--(0,0); 
        \draw[postaction={decorate}] (2,0)--(1,1);
        \draw[postaction={decorate}] (1,1)--(2,2); 
        \node at (-.5,1) { $B$:};
        \node at (0.5,2-.25) {{\small $1$}};
        \node at (0.5+1,2-.25) {{\small $3$}};
        \node at (1.5,.25) {{\small $2$}};
        \node at (1.5-1,.25) {{\small $4$}};
      \end{scope}
      
    \end{scope}

\end{tikzpicture}
  \caption{(Left) A portion of the network. Blue dots stand for the
    multiplication by random phases associated with links, and the red arrows indicate the paths for scattering at a node. (Right) Types of nodes and choice of labeling.}
  \label{fig:network}
\end{figure}
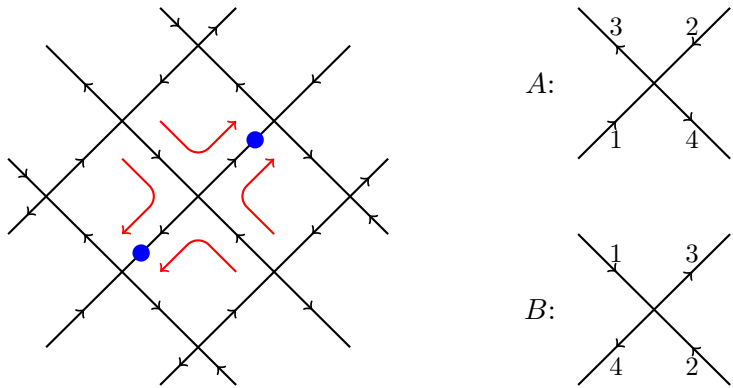

The network model was introduced in \cite{Chalker1988}. It is defined on the directed square lattice sketched in Fig.\ \ref{fig:network}, with two incoming and two outgoing links (or edges) at each node (or vertex). The sets of links and nodes are denoted by ${\cal L}$ and ${\cal N}$ respectively, and we will write $N = |{\cal L}|$ for the number of links. The Hilbert space of the model is
\begin{align}
  \label{eq:H-network}
  {\cal H} = \bigoplus_{\ell \in {\cal L}} \, (\mathbb{C})_\ell = \mathbb{C}^N ,
\end{align}
and its discrete-time evolution operator $U : \; \mathcal{H} \to \mathcal{H}$ is defined as the following $N\times N$ unitary transformation:
\begin{align}
  &\UU = \UUr \, \UUs\, ,
\end{align}
with
\begin{align}
  &\UUr = \bigoplus_{\ell \in {\cal L}} \, \me^{\mathrm{i}\phi_\ell}, \\
  \label{eq:UUsSv}
  &\UUs = \bigoplus_{n \in {\cal N}} \, S_n\, ,
  \quad S_n :\;  (\mathbb{C})_1 \oplus (\mathbb{C})_2
  \to  (\mathbb{C})_3 \oplus  (\mathbb{C})_4 \,.
\end{align}
The unitary numbers $\me^{\mathrm{i} \phi_\ell}$ are phase factors associated to propagation along the links. They are assumed to be independent random variables that are identically distributed according to the uniform measure on the circle. $S_n$ is a $2 \times 2$ unitary transformation connecting incoming to outgoing states at the node $n$, see Fig.\ \ref{fig:network}. Labeling these as $1,2$ and $3,4$ respectively, we take $S_n$ as
\begin{align}
    &S_n \vert 1 \rangle =
    \vert 3 \rangle \, \mathrm{e}^{+\mathrm{i}\pi/4} \cos(t_n) +
    \vert 4 \rangle \, \mathrm{e}^{-\mathrm{i}\pi/4} \sin(t_n) ,
    \label{eq:5-mz} \\
    &S_n \vert 2 \rangle =
    \vert 3 \rangle \, \mathrm{e}^{-\mathrm{i}\pi/4} \sin(t_n) +
    \vert 4 \rangle \, \mathrm{e}^{+\mathrm{i}\pi/4} \cos(t_n) ,
    \label{eq:6-mz}
\end{align}
where each $|\ell\rangle$ ($\ell = 1, \ldots, 4$) is a unit vector in its summand $(\mathbb{C})_\ell \subset {\cal H}$ in \eqref{eq:H-network}. Thus the amplitude for a left turn is $\mathrm{e}^{+\mathrm{i} \pi/4} \cos(t_n)$, while the amplitude for a right turn is $\mathrm{e}^{-\mathrm{i} \pi/4} \sin(t_n)$, and we assume that $0 \leq t_n \leq \pi/2$. The network model is at a critical point when $t_n \equiv t$ (independent of the node $n$) and the probability for turning right equals that for turning left: $\cos^2(t) = \sin^2(t) = 1/2$ \cite{Chalker1988}.

A more general situation is that of an anisotropic network. For this generalization, one thinks of the square network as a bipartite lattice with two types of nodes labeled by, say, $A$ and $B$, and one takes the couplings $t_n$ to depend only on the type of node ($t_n = t_A$ resp.\ $t_n = t_B$). The critical point now occurs when the probability for turning right (left) at a node of type $A$ equals that for turning left (right) at a node of type $B$, i.e., when $\sin(t_A) = \cos(t_B)$.

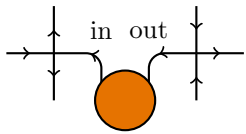
\begin{figure}
  \centering
  \begin{tikzpicture}[thick,decoration={ markings, mark=at position
    0.5 with {\arrow{>}}}, scale=1.25]
  \draw[postaction={decorate}] (0,0)--(0.5,0);
  \draw[postaction={decorate}] (0.5,0)--(0.5,0.5);
  \draw[postaction={decorate}] (0.5,0)--(0.5,-0.5);

  \draw[postaction={decorate}] (2,0.5)--(2,0);
  \draw[postaction={decorate}] (2,-0.5)--(2,0);
  \draw[postaction={decorate}] (2,0)--(2.5,0);

  \draw[postaction={decorate},rounded corners=0.2cm]
  (1,-0.35)-- ++(0,0.35)-- ++(-0.5,0);
  \draw[postaction={decorate},rounded corners=0.2cm]
  (2,0)-- ++(-0.5,0)-- ++(0,-0.35);
  \filldraw[fill=orange!90!black
  ] (1.25,-.5) circle (9pt); 
  \node at (1,.25) {{\small in}};
  \node at (1.5,.25) {{\small out}};
\end{tikzpicture}
  \caption{A point contact. Incoming (in) and outgoing (out)
    semi-links are connected to a charge reservoir.}
  \label{fig:pc}
\end{figure}

The setup described so far is for a closed network. Consider now a set of distinguished links labeled $c_1, \dots, c_r$, which we call point contacts, and form the projectors
\begin{align}\label{eq:PQ}
  P = \sum_{i=1}^r |c_i\rangle\langle c_i| \,, \quad
  Q = 1-P .
\end{align}
Point contacts were introduced in the present context in \cite{Janssen1999}. They are severed links, where both ends that result from cutting the link are connected to charge reservoirs, see Fig.\ \ref{fig:pc}.
The incoming semi-link serves to inject current into the network, while the outgoing semi-link removes current from the network. Thus a point contact acts as a source and as a drain of charge or probability current. The loss of current through outgoing semi-links amounts to a loss of unitarity. This modification of the dynamics is taken into account by concatenating the closed-network time evolution operator $U$ with a factor of $Q$:
\begin{align}
  T = Q \UU .
\end{align}
Thus $T$ is the time-evolution operator of the open network.

\subsection{Scattering states}\label{sect:2.2}

Our interest in this work will be in states of the open network which are stationary (with quasi-energy $E$) and satisfy scattering boundary conditions, i.e., incoming-wave or outgoing-wave boundary conditions at the point contacts. More precisely, we are interested in the \emph{statistical} properties of the wave amplitudes of such states, which we refer to as ``scattering states'' for short. By the choice of a uniform distribution for the random phases $\mathrm{e}^{\mathrm{i} \phi_\ell}$, the wavefunction statistics is the same for all quasi-energies $E$. Thus without loss of generality we set $E = 0$ and look, roughly speaking, for solutions $\vert \psi \rangle$ of the stationarity condition $U \vert \psi \rangle = \vert \psi \rangle$.

To explain this in detail, we first introduce some language and notation. The distinction between links without and with a point contact divides the network into a ``bulk'' and a ``contact'' part. When instituted for the Hilbert space $\mathcal{H}$, this division determines an orthogonal decomposition
\begin{align}
    {\cal H} = \Hbulk \oplus \Hcontact\,, \quad \Hcontact = P\mathcal{H} \simeq \mathbb{C}^r , \quad \Hbulk = Q \mathcal{H} \simeq \mathbb{C}^{N-r} .
\end{align}
To formalize the picture of a point contact being both a source and a drain of current, we define $\Hin$ and $\Hout$ as the spaces of incoming and outgoing states at the point contacts. (Of course we have isomorphisms $\Hin \simeq \Hcontact \simeq \Hout$.) In this way $\UU : \; \mathcal{H} \to \mathcal{H}$ turns into an operator
\begin{align}
    \UU :\; \Hbulk \oplus \Hin \longrightarrow \Hbulk \oplus \Hout\, .
\end{align}
Correspondingly, the forward and backward time-evolution operators of the open network (with loss of probability flux at the point contacts) operate as follows:
\begin{align}
    T = Q\UU &:\; \Hbulk \oplus \Hin \longrightarrow \Hbulk \,, \\
    Q\UU^{-1} &:\; \Hbulk \oplus \Hout \longrightarrow \Hbulk\, .
\end{align}
In order for the dynamics to be free of trivial processes, we assume that the evolution for a single time step does not send back any of the network-incoming flux into the point contacts; in formulas:
\begin{align}
  \label{eq:assumption}
  Q \UU |\psiin\rangle = \UU |\psiin\rangle, \quad
  Q\UU^{-1} |\psiout\rangle = \UU^{-1} |\psiout\rangle ,
\end{align}
where $|\psiin \rangle \in \Hin$ and $|\psiout \rangle \in \Hout$.

\begin{figure}[tb!]
  \centering
    \includegraphics[width=\textwidth]{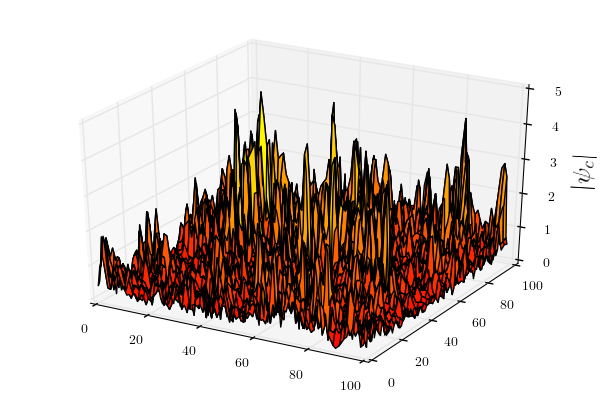}
    \[
    \begin{array}{cc}
    \includegraphics[width=.475\textwidth]{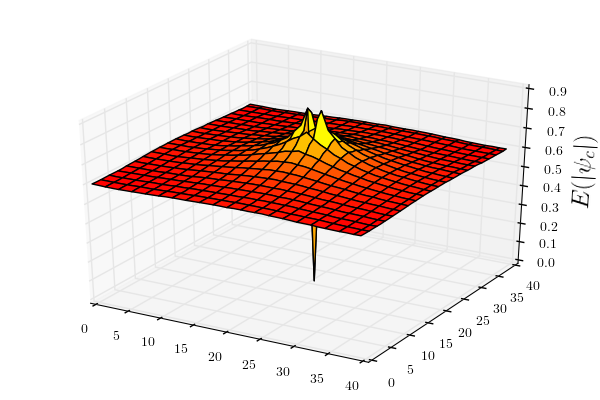}&
    \includegraphics[width=.475\textwidth]{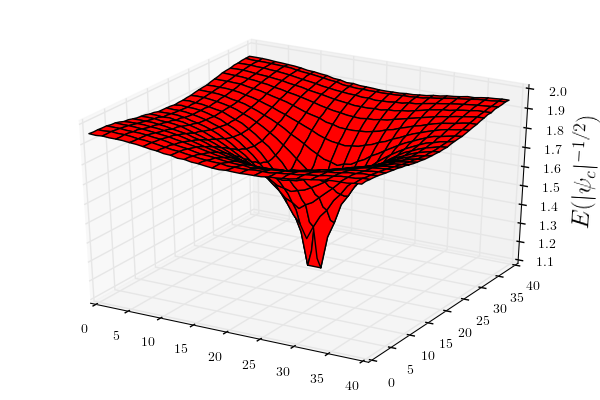}
    \\
      { } & { }
    \end{array}
    \]
    \caption{(Top) The absolute value $|\psi_c|$ of a scattering-state
      wavefunction for a fixed disorder realization. The network has
      $50 \times 50$ plaquettes and periodic boundary conditions
      (pbc). (Bottom) The disorder average of $|\psi_c|$ (left) and
      $|\psi_c|^{-1/2}$ (right) for a network of $20 \times 20$
      plaquettes with pbc. The wavefunction is pinned to zero at the
      center where the network is connected to a point contact. The
      average is over $5 \times 10^5$ disorder realizations. The code used to produce the scattering-state wavefunctions is available at
        \url{https://github.com/rbondesan/network_model/tree/master}
    } \label{fig:pics}
\end{figure}

The quantum-dynamical system evolving a state $|\psi \rangle \in \Hbulk$ from discrete time $t$ to time $t+1$ is then defined by
\begin{align}
  |\psi(0)\rangle = 0 , \qquad
  |\psi(t+1)\rangle = T \big( |\psi(t)\rangle + |\psiin\rangle \big) \qquad (t = 0, 1, 2, \ldots) .
  \label{eq:dynamics}
\end{align}
Here $| \psiin \rangle \in \Hin$ specifies an incoming-wave boundary condition, namely what are the wave amplitudes of the particle flux constantly injected at the point contacts per unit of time. If $N < \infty$ and the dynamics \eqref{eq:dynamics} is run for a very long time, then the state $| \psi(t) \rangle$ will ultimately settle down to a stationary state $| \psi_+ \rangle = \lim_{t \to + \infty} | \psi(t) \rangle$ -- which is what we call a ``scattering state'' for short. An explicit formula for $| \psi_+ \rangle $ can be had by solving the stationarity condition $| \psi(t+1) \rangle = | \psi(t) \rangle$:
\begin{align}
  | \psi_+ \rangle = (1 - T)^{-1} T | \psiin \rangle .
\end{align}
Fig.\ \ref{fig:pics} gives an idea of how such scattering states look at criticality.

Special scattering states are formed by choosing the incoming-wave boundary conditions to inject flux into just a single point contact:
\begin{align}
  |\psi_{+,\, c_i} \rangle = (1-T)^{-1} T | c_i \rangle ,
  \quad i = 1 , \dots , r .
\end{align}
These states are linearly independent and thus span an $r$-dimensional subspace of $\Hbulk$. Note, however, that they are neither orthogonal amongst each other (as vectors in the Hilbert space $\Hbulk$), nor are they normalized to unity:
$\langle \psi_{+,\, c_i} | \psi_{+,\, c_i} \rangle \neq 1$.

Considering the backward dynamics given by $Q\UU^{-1}$, one can construct scattering states $|\psi_- \rangle \in \Hbulk$ that satisfy outgoing-wave boundary conditions. For these the network-outgoing flux is prescribed to be $| \psiout \rangle \in \Hout$:
\begin{align}
    | \psi_- \rangle = Q\UU^{-1} \big( |\psi_- \rangle + | \psiout \rangle \big) .
\end{align}
As before, we define special scattering states for outgoing-wave boundary conditions with one unit of flux on a prescribed outgoing link:
\begin{align}\label{eq:OWST}
  |\psi_{-,\,c_i} \rangle = (1- Q\UU^{-1})^{-1} Q \UU^{-1} |c_i\rangle ,
  \quad i = 1, \dots, r .
\end{align}
We note that the states $| \psi_{- ,\, c_i} \rangle \in \Hbulk$ for outgoing-wave boundary conditions and the states $| \psi_{+ ,\, c_i} \rangle \in \Hbulk$ for incoming-wave boundary conditions span the \emph{same} $r$-dimensional subspace of $\Hbulk$. Indeed, by a foundational and standard result of quantum scattering theory, they are related by a unitary $r \times r$ matrix known as the scattering matrix $S$:
\begin{align}\label{eq:in-out}
    |\psi_{+,\,c_i} \rangle = \sum_{j = 1}^r |\psi_{-,\,c_j} \rangle
    S_{j i} \,.
\end{align}
To verify this relation, one uses $\langle c_j \vert U \psi_{-,\,c_i} \rangle = \langle c_j \vert c_i \rangle = \delta_{ij}$ and the definition of the scattering matrix elements by $S_{ji} = \langle c_j | U \psi_{+,\,c_i} \rangle$.

\subsection{Wavefunction observables}
\label{sect:obs}

From here on, we shall take all point contacts $c_i$ to be located in a small part of the network, $\mathcal{C}$, which we call the ``contact region''. Our interest is in the statistics of wave intensities of scattering states as a function of the distance from that region. With this motivation, we will now describe a large class of scattering-state observables that are informative and tractable by the methods of the present paper.

In order to do quantitative analysis (especially, to have coarse-grained observables), we choose a positive measure $\ell \mapsto w(\ell) \geq 0$ supported on a small region $\mathcal{R}$, which we call an ``observation region''. Assuming $r$ point contacts, we form the $r \times r$ matrix $K(w)$ with matrix elements
\begin{align}
    K_{ij}(w) = \sum_{\ell \in \mathcal{R}}
    w(\ell) \bar\psi_{c_i} (\ell) \psi_{c_j} (\ell) ,
\end{align}
where $\psi_{c_i}(\ell) \equiv \langle \ell \vert \psi_{+,c_i} \rangle$ ($i = 1, \ldots, r$) are the wave amplitudes of the scattering states of Sect.\ \ref{sect:2.2} for incoming-wave boundary conditions (or outgoing-wave boundary conditions; it doesn't matter). Note that the matrix $K(w) \equiv K$ is Hermitian and positive-definite.

Our observables of interest are formed by taking the $r \times r$ matrix $K = K(w)$ to be the argument of a Schur polynomial (or generalizations thereof). Recall that Schur poly\-nomials are defined as the characters of polynomial irreducible representations of $\mathrm{GL}(r)$, and the most general such representation is labeled by $r$ integers $\lambda_1 \geq \lambda_2 \geq \ldots \geq \lambda_r \geq 0$. Let $\lambda \equiv (\lambda_1, \ldots, \lambda_r)$ for short. Then if $k_1, \ldots, k_r$ are the eigenvalues of $K$, the Schur polynomial $s_\lambda(K)$ can be expressed as a ratio of determinants:
\begin{align}\label{eq:schur}
    s_\lambda(K) = \Delta(k_1,\ldots,k_r)^{-1} \Det \begin{pmatrix}
    k_1^{\lambda_1 +r-1} &k_2^{\lambda_1 +r-1} &\ldots &k_r^{\lambda_1 +r-1}\\
    k_1^{\lambda_2 +r-2} &k_2^{\lambda_2 +r-2} &\ldots &k_r^{\lambda_2 +r-2}\\
    \vdots & \vdots & \ddots & \vdots \\
    k_1^{\lambda_r} &k_2^{\lambda_r} &\ldots &k_r^{\lambda_r}\\
    \end{pmatrix} ,
\end{align}
with $\Delta$ the Vandermonde determinant:
\begin{align}
  \Delta(k_1, \dots, k_r) = \prod_{i<j} (k_i - k_j) .
\end{align}
For $\lambda_1 = \ldots = \lambda_p = 1$ and $\lambda_{p+1} = \ldots = \lambda_r = 0$ (physically speaking, $p$ fermions on $\mathbb{C}^r$) one gets the elementary symmetric polynomials
\begin{align}
    s_\lambda(K) \equiv e_p(K) = \sum_{1 \leq i_1 < \ldots < i_p \leq r} k_{i_1} k_{i_2} \cdots k_{i_p}\, .
\end{align}
Their invariant expressions in terms of traces are given by the
Newton-Girard formulas:
\begin{align}
  e_1 = p_1\,, \quad e_2 = \tfrac{1}{2}p_1^2 -\tfrac{1}{2}p_2 \,, \quad
  e_3 = \tfrac{1}{6}p_1^3 - \tfrac{1}{2}p_1p_2+\tfrac{1}{3}p_3 \,, \quad
  \text{etc.}, \quad p_n &\equiv\mathrm{Tr}(K^n) .
\end{align}
Another Schur polynomial of special interest is the complete homogeneous symmetric polynomial $h_q\,$, given by $\lambda_1 = q$, $\lambda_2 = \ldots = \lambda_r = 0$. In this case ($q$ bosons on $\mathbb{C}^r$) one has
\begin{align}
    s_\lambda(K) \equiv h_q(K) = \sum_{1 \leq i_1 \leq \ldots \leq i_q \leq r} k_{i_1} k_{i_2} \cdots k_{i_q} ,
\end{align}
or, in invariant terms,
\begin{align}
  h_1 = p_1 \,, \quad
  h_2 = \tfrac{1}{2}p_1^2 +\tfrac{1}{2}p_2 \,, \quad
  h_3 = \tfrac{1}{6}p_1^3 + \tfrac{1}{2}p_1p_2+\tfrac{1}{3}p_3 \,,
  \quad \text{etc.}
\end{align}

A very important remark is that, since all eigenvalues $k_1, \ldots, k_r$ of our matrix $K$ are positive, the Schur polynomials $s_\lambda(K)$ also make sense for \emph{complex} values of $\lambda_1, \ldots, \lambda_r\,$, by the determinantal expression \eqref{eq:schur} (although no longer as characters of polynomial representations). Our most general local observable then is (the disorder average) of $s_\lambda(K)$ for a set of complex numbers $\lambda = (\lambda_1, \ldots, \lambda_r)$. Let us also remark that by Eq.\ \eqref{eq:in-out} all functions $s_\lambda(K)$ indeed are independent of the choice of incoming-wave or outgoing-wave boundary conditions for the scattering states.

The precise nature of the wavefunction observables $s_\lambda(K)$ is best understood by looking at some examples. In the special case of $r = 1$ with a single contact $c$ and a Dirac measure $w_{\ell}(\ell^\prime) = \delta_{\ell \ell^\prime}$ one has
\begin{align}
    h_q\big( K(w_{\ell}) \big) = | \psi_{c}(\ell) |^{2 q} .
\end{align}
This is the observable to be studied in much detail in Sect.\ \ref{sec:6}, where $q$ will be allowed to be complex. In the special case of $r$ contacts and a Dirac measure $w^{(r)}(\ell) = \sum_{i=1}^r \delta_{\ell_i\, \ell}$ supported on $r$ observation links one finds
\begin{align}
    e_r \big( K(w^{(r)}) \big) = \left\vert \mathrm{Det} \big( \psi_{c_i} (\ell_{i^\prime}) \big)_{i,i^\prime = 1,\ldots,r} \right\vert^2 .
\end{align}

More generally, we can have several disjoint observation regions $\mathcal{R}_1, \ldots, \mathcal{R}_n$ equipped with positive measures $w_1, \ldots, w_n$ where $w_j$ is supported in $\mathcal{R}_j$ ($j = 1, \ldots, n$). Each observation region $\mathcal{R}_j$ is assumed to be ``small'' (from the coarse-grained perspective of a scaling limit) and can therefore be assigned a well-defined position $x_j$ in a suitably coarse-grained network. The matrix $K$ then consists of $n$ local pieces:
\begin{align}\label{eq:30-mz}
    K = \sum_{j=1}^n K(x_j) , \quad K_{i i^\prime}(x_j) = \sum_{\ell \in \mathcal{R}_j} w_j(\ell)\, \bar\psi_{c_i} (\ell) \psi_{c_{i^\prime}} (\ell) .
\end{align}
Our most general observable will be a disorder-averaged multi-point correlation function of the local wavefunction data $K(x_1), \ldots, K(x_n)$. We refer to Sect.\ \ref{sect:5.2} for the details.

\subsection{Observables from Green functions}

We now cast our multi-point correlation functions into a form suitable for processing and evaluation in the SUSY vertex model of Sect.\ \ref{sect:4}.

Our first observation is that the trace of any power $m \in \mathbb{N}$ of the $r \times r$-matrix $K$ can be re-expressed as
\begin{align}\label{eq:fromK2G}
    \mathrm{Tr}_{\mathbb{C}^r} (K^m) = \mathrm{Tr}_\mathcal{\Hbulk} (G \cdot w)^m
\end{align}
by the cyclic property of the trace operation. Here $w = \sum_\ell \vert \ell \rangle w(\ell) \langle \ell \vert$ is the link-diagonal operator associated with the positive measure $w$, and the operator $G : \; \Hbulk \to \Hbulk$ (a kind of Green operator) is defined in terms of the scattering states $\vert \psi_{\pm,\, c_i} \rangle$ by
\begin{align}\label{eq:Green-op}
    G = \sum_{i=1}^r |\psi_{+,\,c_i}\rangle \langle \psi_{+,\,c_i}| = \sum_{i=1}^r |\psi_{-,\,c_i}\rangle \langle \psi_{-,\,c_i}| .
\end{align}
Like $K$, this operator $G$ is self-adjoint and positive-definite of rank $r$.

Recalling the fact that all Schur functions $s_\lambda$ are determined by traces, it follows from the relation \eqref{eq:fromK2G} that
\begin{align}
    s_\lambda \big( K(w) \big) = s_\lambda(G \cdot w) .
\end{align}
Thus we can express all observables of Sect.\ \ref{sect:obs} as functions of $G \cdot w$. To make progress with the analytical theory of these observables, we need a good formula for $G$, making it calculable by the supersymmetry method of Sect.\ \ref{ref:2nd}. Our first innovation in the present paper is the derivation of such a formula, as follows.

By recalling the definition of the projector $P$ in Eq.\ \eqref{eq:PQ} and the expression \eqref{eq:OWST} for the scattering states with outgoing-wave boundary conditions, we can write $G$ as
\begin{align}
    G = \sum_i |\psi_{-,\,c_i}\rangle\langle \psi_{-,\,c_i}|
    = Q (1 - T^\dagger)^{-1} \UU^{-1} P \UU (1 - T)^{-1} Q ,
\end{align}
where $T^\dagger = U^{-1} Q$ is the adjoint of $T = Q U$. Now, the middle factor can be reorganized as
\begin{align}
    \UU^{-1} P \UU = 1 - T^\dagger T = (1 - T^\dagger)(1 - T) + (1 - T^\dagger) T + T^\dagger (1 - T) ,
\end{align}
and hence
\begin{align}
    G = Q + T (1 - T)^{-1} Q + Q (1 - T^\dagger)^{-1} T^\dagger .
\end{align}
When taking matrix elements between bulk states we obtain
\begin{align}\label{eq:MEofK}
    \langle \ell \vert G \vert \ell^\prime \rangle = \langle \ell \vert T (1 - T)^{-1} + (1 - T^\dagger)^{-1} \vert \ell^\prime \rangle ,
\end{align}
since $Q + Q (1-T^\dagger)^{-1} T^\dagger = Q (1-T^\dagger)^{-1}$. This is the desired formula for the Green operator $G$, which will serve as our reference point in Sect.\ \ref{sect:5.2}.

For later reference, let us also note the identity
\begin{align}
    \frac{T}{1-T} + \frac{1}{1-T^\dagger} = \frac{1}{2} \left( \frac{1+T}{1-T} + \frac{1+T^\dagger}{1 - T^\dagger} \right) .
\end{align}
Hence, defining the real part of an operator $A$ as $\mathrm{Re}\, A = \frac{1}{2} (A + A^\dagger)$, one has $G = \mathrm{Re} (Q A_T Q)$ with $A_T = (1+T) (1-T)^{-1}$ the Cauchy transform of $T$.

As a word of caution, we should mention that the inverse of $1-T$ as an operator on the full Hilbert space $\mathcal{H} = \Hbulk \oplus \Hcontact$ is not well-defined when there is an eigenvector $|\phi \rangle$ of $\UU$ which has eigenvalue $1$ and also satisfies $Q | \phi\rangle = |\phi \rangle$. This happens when the amplitudes for the two incoming links at a node conspire to make an outgoing amplitude vanish, so that placing a point contact on that outgoing link does not modify the wavefunction. However, this scenario (called a ``bound state in the continuum'') does not pose a problem in our scattering-theoretic setting. Indeed, any such state $| \phi \rangle$ satisfies $\langle \phi | \psiin \rangle = \langle \phi | Q | \psiin \rangle = 0$, and from the orthogonality $\langle \phi | \psi(t) \rangle = 0$ at time $t$ one infers the orthogonality $\langle \phi | \psi(t+1) \rangle = 0$ at time $t+1$ by the dynamics of Eq.\ \eqref{eq:dynamics}. Thus by induction on $t$ it follows that the bound state $| \phi \rangle$ is orthogonal to the stationary limit $\lim_{t \to \infty} | \psi(t) \rangle$ of the scattering state.


\section{Second quantization of the network model}
\label{ref:2nd}

To perform the disorder average of our network-model observables, we employ a variant of the Wegner-Efetov supersymmetry method tailored to the model at hand. This variant was originally conceived by N.\ Read (unpublished notes) and has already appeared in several works in the past (e.g.\ {\cite{Gruzberg1997, Zirnbauer1997}}). The relevant mathematics underlying the formalism was spelled out at length in \cite{Conrey2005}. Nonetheless, to make the present paper self-contained we review the general formalism in Sects.\ \ref{sec:fermions}--\ref{sec:gaussian}, having in mind especially the observables put on stage in Sect.\ \ref{sect:obs}. The application to the network model is described in Sect.\ \ref{sec:2nd_quant}.

The first step is to second-quantize the network-model operators. For this purpose one introduces annihilation and creation operators for retarded ($+$) and advanced ($-$) bosons and fermions at every link $\ell$ of the lattice. Denoted by $b_\pm (\ell)$, $b_\pm^\dagger (\ell)$ and $f_\pm (\ell)$, $f_\pm^\dagger (\ell)$, these satisfy canonical commutation and anti-commutation relations:
\begin{align}
    &[ b_\alpha(\ell) , b_\beta^\dagger(\ell')] = \delta_{\alpha\beta}\, \delta_{\ell \ell'} , \qquad [ b_\alpha(\ell) , b_\beta(\ell') ] = [ b_\alpha^\dagger(\ell) , b_\beta^\dagger(\ell') ] = 0 ,
    \label{eq:CCR} \\
    &\{f_\alpha(\ell), f_\beta^\dagger (\ell') \} = \delta_{\alpha\beta}\, \delta_{\ell \ell'} , \qquad \{ f_\alpha(\ell) , f_\beta(\ell')\} = \{f_\alpha^\dagger(\ell), f_\beta^\dagger(\ell')\} = 0, \label{eq:CAR}
\end{align}
for all $\alpha,\beta = \pm$ and all links $\ell, \ell'\in {\cal L}$ of the network. We write ${\cal F}$ for the Fock space generated by the boson and fermion creation operators acting on the particle vacuum.

In the following, we apply the standard procedure of second quantization to the network-model operators, emphasizing the aspects of relevance for our goal.

\subsection{Fermionic formulas}
\label{sec:fermions}

Given an invertible operator $g = \me^X$ on the Hilbert space $\mathcal{H}$ of the network model, we define two operators $\sigma_\pm(g)$ acting on the Fock space $\mathcal{F}$ by
\begin{align}
    \sigma_+\left(\me^X \right) &=
    \exp \sum_{\ell,\ell' \in {\cal L}}f_+^\dagger(\ell) \langle \ell \vert X \vert \ell' \rangle f_+(\ell') \equiv
    \exp f^\dagger_+ X f_+ \,, \\
    \sigma_-\left(\me^X \right) &=
    \exp \sum_{\ell,\ell' \in {\cal L}} f_-(\ell) \langle \ell \vert X \vert \ell' \rangle f_-^\dagger (\ell') \equiv
    \exp f_- X f_-^\dagger \,.
\end{align}
By letting $g$ vary we get two maps $\sigma_\pm : \; \GL({\cal H}) \to \GL({\cal F})$. According to the principles of second quantization, these maps are Lie group homomorphisms (or representations), i.e., they satisfy $\sigma_\pm(g h) = \sigma_\pm (g) \sigma_\pm (h)$. (This standard fact will be reviewed in some detail below.) We refer to $\sigma_\pm$ as the retarded and advanced representations. They are related to each other by a particle-hole transformation exchanging creation with annihilation operators:
\begin{align}\label{eq:rel+-}
    \sigma_-(g) = \Det(g)\, \sigma_+ (g^T)^{-1} \Big|_{f_+ \to f_-} .
\end{align}
Their infinitesimal versions are maps $\dd \sigma_{\pm} :\; \gl({\cal H}) \to \gl({\cal F})$ defined by
\begin{align}
  \label{eq:sigma*+}
  &\dd\sigma_+ (X) = \frac{\dd}{\dd t} \sigma_+(\me^{tX}) \Big\vert_{t=0}
  = f^\dagger_+ X f_+ \,, \\
  &\dd\sigma_- (X) = \frac{\dd}{\dd t} \sigma_-(\me^{tX}) \Big\vert_{t=0}
  = f_- X f^\dagger_- \,.
\end{align}
{}From the canonical anti-commutation relations \eqref{eq:CAR}, it is easy to see that both $\dd \sigma_+$ and $\dd \sigma_-$ are Lie algebra representations:
\begin{align}\label{eq:32mz}
  [\dd\sigma_\pm (X), \dd\sigma_\pm (Y)] = \dd\sigma_\pm ([X,Y]) .
\end{align}
Taking the commutator with the fermion Fock operators we have
\begin{align}
    [\dd\sigma_+ (X) , f^\dagger_+] =  f^\dagger_+ X ,
    \qquad [\dd\sigma_+(X) , f_+] = - X f_+ \,, \\
    [\dd\sigma_- (X) , f_-] = f_- X ,\qquad
    [\dd\sigma_- (X), f^\dagger_-] = - X f_-^\dagger \,.
\end{align}
By exponentiating this commutator action and using the identity $\me^{A} B \, \me^{-A} = \me^{[A , \,\cdot]} B$, one obtains
\begin{align}
  \label{eq:Adsigma}
  \sigma_+(g)
  \begin{pmatrix}
    f^\dagger_+ & f_+
  \end{pmatrix}
  \sigma_+(g)^{-1}
  =
  \begin{pmatrix}
    f^\dagger_+ g & g^{-1} f_+
  \end{pmatrix} , \\
  \sigma_-(g)
  \begin{pmatrix}
     f_- & f^\dagger_-
  \end{pmatrix}
  \sigma_-(g)^{-1}
  =
  \begin{pmatrix}
    f_- g & g^{-1} f^\dagger_-
  \end{pmatrix} .
\end{align}
Thus $f^\dagger_+(\ell)$ and $f_-(\ell)$ transform in the same way as the basis vectors $\vert \ell \rangle$ of $\mathcal{H}$ do under the fundamental action $g \vert \ell \rangle = \sum \vert \ell^\prime \rangle \langle \ell^\prime \vert g \vert \ell \rangle$ of $\GL({\cal H})$ on $\mathcal{H}$, while $f_+(\ell)$ and $f^\dagger_-(\ell)$ transform like co-vectors in the anti-fundamental representation.

Useful objects to consider are the (super-)traces over Fock space of the two representations $\sigma_\pm (g)$. The results for these traces are quite simple (and motivate why the second-quantized representation is introduced in the first place):
\begin{align}
    \STr \sigma_+(g) &= \Det(1-g) , \label{eq:STrDet+} \\
    \STr \sigma_-(g) &= \Det(g)\, \Det(1-{g^{-1}}^T) = (-1)^N \Det(1-g) .
    \label{eq:STrDet-}
\end{align}
In both cases, the supertrace $\STr$ is understood to be over the respective Fock space. It is defined by $\STr \sigma_\pm(g) = \Tr \, (-1)^{n_f} \sigma_\pm(g) $ where $n_f$ is the number of fermions.

These formulas are immediate if $g$ is diagonal in the chosen link basis of $\mathcal{H}$. For diagonalizable $g = u \lambda u^{-1}$ they follow by the conjugation invariance of the determinant, the representation property $\sigma_\pm(u \lambda u^{-1}) = \sigma_\pm(u)\, \sigma_\pm(\lambda)\, \sigma_\pm(u)^{-1}$ and the invariance of the trace under cyclic permutations. The case of non-diagonalizable $g$ then follows by analytic extension. There also exists a direct proof for the general case using some standard multi-linear algebra not recorded here. Note also that the second formula follows from the first one by the relation \eqref{eq:rel+-}.

\subsection{Bosonic formulas}

We now turn to the bosonic variants of the formulas above. The final formulas look very similar to those for fermions, but the analysis here needs a little more care. For $X \in \gl({\cal H})$ we put
\begin{align}
  \dd \omega_+ (X) &=
  \sum_{\ell,\ell' \in {\cal L}} b^\dagger_+ (\ell)
  \langle \ell \vert X \vert \ell' \rangle b_+(\ell')
  \equiv
  b ^\dagger_+ X b_+ \,, \label{eq:39mz}
  \\
  \dd \omega_- (X) &=
  - \sum_{\ell,\ell' \in {\cal L}} b_- (\ell)
  \langle \ell \vert X \vert \ell' \rangle b^\dagger_-(\ell')
  \equiv
  - b_- X b_-^\dagger \,. \label{eq:40mz}
\end{align}
Similar to the case \eqref{eq:32mz} of fermions, these maps $\dd \omega_\pm :\; \gl({\cal H}) \to \gl({\cal F})$ are easily seen to be homomorphisms of the commutator:
\begin{align}
  [\dd \omega_\pm (X), \dd \omega_\pm (Y)] = \dd \omega_\pm ([X,Y]) .
\end{align}
Thus they are Lie algebra representations, and our notation $\dd \omega_\pm$ indicates that we are going to think of them as the infinitesimal versions of two Lie group representations $\omega_\pm$ as before. The additional minus sign in the advanced case reflects the fact that $b \mapsto b^\dagger$ and $b^\dagger \mapsto -b$ (as opposed to $f \mapsto f^\dagger$ and $f^\dagger \mapsto + f$ for fermions) is an automorphism of the operator algebra for bosons.

The Lie algebra elements $\dd \omega_\pm(X)$ act on the Fock operators $b, b^\dagger$ by the commutator:
\begin{align}
  [\dd \omega_+ (X), b^\dagger_+] =  b_+^\dagger X , \qquad
  [\dd \omega_+ (X), b_+] = -X b_+ \,,\\
  [\dd \omega_- (X), b_-] = b_-X ,\qquad
  [\dd \omega_- (X), b_-^\dagger] = - X b^\dagger_- \,.
\end{align}
We observe that these formulas look exactly the same as their fermionic counterparts.

Next we exponentiate the operators introduced in (\ref{eq:39mz}, \ref{eq:40mz}) to define
\begin{align}
    \omega_\pm (\me^X) = \exp \left( \dd \omega_\pm (X) \right) .
\end{align}
Each of the two resulting maps $\omega_\pm :\; \GL({\cal H}) \to \GL({\cal F})$ still has the representation property $\omega_\pm (g h) = \omega_\pm (g)\, \omega_\pm (h)$. This prompts a comment. In general, a Lie algebra representation on the bosonic Fock space does \emph{not} exponentiate to a Lie group representation on Fock space. (The difficulty is that the bosonic Fock space is infinite-dimensional and one has to restrict to a semi-group of contractions in order for the exponentiated operators to exist. This will happen in Sect.\ \ref{sect:5.1} below.) However, in the present instance we have a simplification because the Lie algebra representations (\ref{eq:39mz}, \ref{eq:40mz}) conserve the number of bosons. Thus the infinite-dimensional representations $\omega_\pm$ are really just (infinite) sums of finite-dimensional representations, one for each boson number, and on every single one of these the operators $\omega_\pm(g)$ make sense for any $g \in \GL({\cal H})$.

As before, there exists a relation
\begin{align}
  \omega_- (g^{-1}) = \Det(g)\, \omega_+ (g^T) \Big\vert_{b_+\to b_-}\,,
\end{align}
and we have an action on boson operators by conjugation:
\begin{align}
  \label{eq:Adomega}
  \omega_+(g)
  \begin{pmatrix}
    b^\dagger_+ & b_+
  \end{pmatrix}
  \omega_+(g)^{-1}
  &=
  \begin{pmatrix}
    b^\dagger_+ g & g^{-1} b_+
  \end{pmatrix} ,
  \\
  \omega_-(g)
  \begin{pmatrix}
    b_- & b^\dagger_-
  \end{pmatrix}
  \omega_-(g)^{-1}
  &=
  \begin{pmatrix}
    b_- g & g^{-1} b^\dagger_-
  \end{pmatrix} .
\end{align}

Now the operator $\omega_+(g)$ on $\mathcal{F}$ is trace class for $g$ in the semi-group in $\GL({\cal H})$ of contractions ($g^\dagger g < 1$), while $\omega_-(g)$ is trace class for $g$ in the semi-group of expansions ($g^\dagger g > 1$). Thus in both cases the trace over Fock space exists as an absolutely convergent sum. The outcomes are reciprocal to the corresponding expressions for fermions:
\begin{align}
  \Tr \omega_+(g) &= \Det^{-1}(1-g), \label{eq:48mz} \\
  \Tr \omega_-(g) &= \Det^{-1}(g) \Det^{-1} \left( 1 - {g^{-1}}^T \right)
  = (-1)^N \Det^{-1}(1-g) . \label{eq:49mz}
\end{align}
Again, these formulas are immediate if $g$ is diagonal (the trace then factors and one just has to sum a geometric series for each eigenvalue or link $\ell \in {\cal L}$). The general case requires more mathematical effort; see {\cite{Howe1988, Conrey2005}} and \cite{HPZ}.

Let us anticipate that we are going to apply the formulas (\ref{eq:48mz}, \ref{eq:49mz}) in the following way: we will take the trace of $g = Q_\epsilon U$ in the retarded representation $\omega_+$ and the trace of $g = Q_\epsilon^{-1} U$ in the advanced representation $\omega_-\,$, where $Q_\epsilon = \mathrm{e}^{-\epsilon} Q + (1-\mathrm{e}^{-\epsilon})P$ and $\epsilon > 0$ is a regularization parameter that will ultimately be removed ($\epsilon \to 0+$).

\subsection{Gaussian integral representations}
\label{sec:gaussian}

We have claimed that the Lie algebra representations $\dd \sigma_\pm$ exponentiate to group representations, and so do the $\dd \omega_\pm$ under the restriction to suitable semi-groups in $\GL({\cal H})$. To substantiate this claim, we follow R.\ Howe \cite{Howe1988} and consider an alternative representation by Gauss--Berezin integral operators, which also has the advantage of being more manageable for some computational purposes. The bosonic version can be found in \cite{Conrey2005}; see also \cite{HPZ}. For the fermionic and super-variants we do not know of a good reference but for \cite{Neretin}.

We start by discussing the fermionic version, focusing on $\sigma_+$ first. To keep the notation simple, we write $\sigma_+ \equiv \sigma$ for now; the omitted index will be re-instated in due course. Introducing a Grassmann algebra generated by $2N$ anti-commuting variables $\{ \xi (\ell) , \bar{\xi} (\ell) \}_{\ell \in {\cal L}}\,$, we define the operator
\begin{align}
  \TT_\xi = \me^{ \bar\xi f + \xi f^\dagger} \equiv
  \exp \sum_{\ell\in{\cal L}} \left(
  \bar\xi (\ell) f (\ell) + \xi(\ell) f^\dagger (\ell) \right) .
\end{align}
We use the standard convention that the Grassmann variables $\xi$, $\bar\xi$ anti-commute with the Fock operators $f$, $f^\dagger$. Note that one can rewrite $\TT_\xi$ in normal-ordered fashion:
\begin{align}\label{eq:norm-ord}
  \TT_\xi = \me^{\xi f^\dagger} \me^{\bar{\xi} f} \me^{-\frac{1}{2} \bar{\xi} \xi} .
\end{align}
A property of special importance is the composition law
\begin{align}
  \TT_{\xi} \TT_{\eta} = \TT_{\xi + \eta} \, \me^{-\frac{1}{2} (\bar\xi \eta - \bar\eta \xi)} ,
\end{align}
where $\{ \eta (\ell) , \bar\eta (\ell) \}_{\ell \in {\cal L}}$ are another set of Grassmann algebra generators.

For an element $g \in \GL({\cal H})$ with $1-g$ invertible, consider now the Cauchy map
\begin{align}
  g \mapsto A_g = \frac{1+g}{1-g} \,,
\end{align}
which sends $g$ to another operator $A_g$ on ${\cal H}$. Note the relation $A_{g^{-1}} = - A_g\,$. We then introduce the following operator on Fock space:
\begin{align}\label{eq:int-repn}
  \widetilde{\sigma} (g) = \Det(1-g) \int_\xi \me^{-\frac{1}{2}\bar\xi A_g \xi} \, \TT_\xi \,,
\end{align}
where the integral sign stands for the Berezin integral over anti-commuting variables:
\begin{align}
  \int_{\xi} = \prod_\ell \frac{\partial^2}{\partial\bar{\xi} (\ell)\, \partial\xi(\ell)} \,.
\end{align}

We will now argue that $\widetilde{\sigma} = \sigma_+\,$. For this we first show that $\widetilde{\sigma}$ and $\sigma_+$ agree at the infinitesimal level. To that end, we set $g = 1 + t X$ with invertible $X$ and adopt the normal-ordered form \eqref{eq:norm-ord} of $\TT_\xi\,$. Using the relation $- \frac{1}{2} (A_g +1) = (tX)^{-1}$ we then get
\begin{align}
  \widetilde{\sigma} (1 + t X) &= \Det(- t X) \int_{\xi} \me^{\bar{\xi} (t X)^{-1}\xi}\,\me^{\xi f^\dagger} \, \me^{\bar{\xi} f} \\
  &= \Det(-X) \int_\xi
  \me^{\bar\xi X^{-1}\xi} \, \me^{t\xi f^\dagger} \, \me^{\bar\xi f}\,,
\end{align}
where in the second line we scaled the variables $\xi \to t \xi$.
Now we take the derivative with respect to $t$, set $t=0$, and perform
the Gaussian integral to arrive at
\begin{align}
  \dd \widetilde{\sigma} (X) &= \frac{\dd}{\dd t} \widetilde{\sigma}(1+tX) \Big\vert_{t=0} \\
  &= \Det(-X) \int_{\xi} \me^{\bar{\xi} X^{-1}\xi} \, (\xi f^\dagger)(\bar\xi f) = f^\dagger X f .
\end{align}
Thus $\dd \widetilde{\sigma}$ indeed coincides with $\dd \sigma_+$ of
Eq.\ \eqref{eq:sigma*+}.

In order to extend the equality $\dd \widetilde{\sigma} = \dd \sigma_+$ beyond the infinitesimal level, recall from basic theory that a Lie algebra representation on a finite-dimensional vector space integrates to a Lie group representation by exponentiation if the Lie group is simply connected. (In the absence of the latter property, the exponential map may fail to give a single-valued representation.) Now $\GL({\cal H})$ is not simply connected, as its deformation retract $\mathrm{U}({\cal H})$ is not. Nevertheless, exponentiation of $\dd \sigma_+$ does yield a good (meaning single-valued) integrated representation, since the weights of the Lie algebra representation $\dd \sigma_+$ on Fock space are fermion occupation numbers and hence integral.

Thus the Lie algebra representation $\dd \sigma_+ = \dd \widetilde{\sigma}$ does integrate to a Lie group representation $\sigma_+\,$. Since there cannot be more than one integrated representation, the equality $\sigma_+ = \widetilde{\sigma}$ will follow if we can show that $\widetilde{\sigma}$ has the representation property $\widetilde{\sigma} (g)\, \widetilde{\sigma} (h) = \widetilde{\sigma} (gh)$. Let us therefore demonstrate this property.

Initially we have to assume that $g, h$ are such that all of $\Det(1-g)$, $\Det(1-h)$ and $\Det(1-gh)$ are non-zero, in order for $\widetilde{\sigma} (g)$, $\widetilde{\sigma} (h)$ and $\widetilde{\sigma} (gh)$ to be defined by Eq.\ \eqref{eq:int-repn}. We then use $\Det(1-g) = \Det^{-1} \left(\tfrac{1}{2}(A_g + 1) \right)$, etc., and rewrite the product
\begin{align}
  \widetilde{\sigma}(g)\, \widetilde{\sigma}(h) =
  \Det^{-1}\left(\tfrac{1}{2}(A_g+1)\right)
  \Det^{-1}\left(\tfrac{1}{2}(A_h+1) \right)\\\times
  \int_\xi   \me^{-\frac{1}{2}\bar\xi A_g \xi}\,
  \int_\eta   \me^{-\frac{1}{2}\bar\eta A_h \eta}\,
  \me^{-\frac{1}{2} (\bar\xi\,\eta - \bar\eta \xi)}\,
  \TT_{\xi +\eta}
\end{align}
by changing variables to $\alpha = \xi + \eta$, $\bar\alpha = \bar\xi
+ \bar\eta$ and $\beta = \eta$, $\bar\beta = \bar\eta\,$. The quadratic form in the exponential transforms as
\begin{align}
  &-\frac{1}{2}\bar\xi A_g \xi -
  \frac{1}{2}\bar\eta A_h \eta
  -\frac{1}{2} (\bar\xi\, \eta - \bar\eta \xi) \\
  = &-\frac{1}{2}\bar\alpha A_g \alpha -
  \frac{1}{2}\bar\beta (A_h+A_g) \beta
  + \frac{1}{2} \bar\beta (A_g+1) \alpha
  + \frac{1}{2} \bar\alpha (A_g-1)\beta \,.
\end{align}
Doing then the Gaussian integral over $\beta, \bar\beta$ by completing the square we obtain
\begin{align}
  \widetilde{\sigma}(g)\, \widetilde{\sigma}(h) =
  \Det^{-1}\left(\tfrac{1}{2}(A_g+1)\right)
  \Det^{-1}\left(\tfrac{1}{2}(A_h+1)\right)
  \Det\left(\tfrac{1}{2}(A_g+A_h)\right)\\
  \times \int_\alpha \me^{-\frac{1}{2}\bar{\alpha} \left( A_g-(A_g-1)(A_g+A_h)^{-1}(A_g+1) \right) \alpha}
  \, \TT_{\alpha}  \,.
\end{align}
To prove that this expression for $\widetilde{\sigma}(g)\, \widetilde{\sigma}(h)$ is equal to that for $\widetilde{\sigma}(gh)$, we first note that the quadratic form in the exponent can be written as
\begin{align}
  A_g-(A_g-1)(A_g+A_h)^{-1}(A_g+1) =
  (A_h+1)(A_g+A_h)^{-1}(A_g+1)-1 .
\end{align}
Using simple algebraic manipulations, we then show that this expression equals $A_{gh}$:
\begin{align}
  &(A_h+1)(A_g+A_h)^{-1}(A_g+1) \\
  &= \left( (A_g+1)^{-1}+(A_h+1)^{-1}-2(A_g+1)^{-1}(A_h+1)^{-1} \right)^{-1}\\
  &= 2\big( (1-g)+(1-h)-(1-h)(1-g) \big)^{-1}\\
  \label{eq:Agh}
  &= 2(1-gh)^{-1} = A_{gh} + 1 .
\end{align}
With the help of this equation one can also check that the product of determinants
\begin{align}
  &\Det^{-1}\left(\tfrac{1}{2}(A_h+1)\right)
  \Det^{-1}\left(\tfrac{1}{2}(A_g+1)\right)
  \Det\left(\tfrac{1}{2}(A_g+A_h)\right)
  \\
  = &\Det^{-1}
  \left(
    \tfrac{1}{2}(A_h+1)
    (A_g+A_h)^{-1}
    (A_g+1)
  \right)
\end{align}
reduces to the one appearing in $\widetilde{\sigma}(gh)$, namely $\Det^{-1} \left( \tfrac{1}{2} (A_{gh}+1) \right)$. This proves that $\widetilde{\sigma}(g)\, \widetilde{\sigma}(h) = \widetilde{\sigma} (gh)$. It also completes the proof that the operators $\widetilde{\sigma}(g)$ and $\sigma_+ (g)$ are the same when both are defined. Moreover, it allows us to extend $\widetilde{\sigma}_+$, initially defined only for $g \in \GL({\cal H})$ with $1-g$ invertible, to all $g\in \GL({\cal H})$. In the following we will abandon the distinction between $\sigma_+$ and $\widetilde{\sigma}$ and refer to both as $\sigma_+\,$.

After this extensive discussion of $\sigma_+$, let us turn to $\sigma_-\,$. Given $\sigma_+\,$, a quick way of arriving at the Gauss--Berezin integral formula for $\sigma_-(g)$ is to use Eq.\ \eqref{eq:rel+-} and the relation $A_{(g^{-1})^T} = - (A_g)^T$. The resulting expression is
\begin{align}
  \sigma_-(g) = \Det(1-g) \int_\xi \me^{-\frac{1}{2}\bar\xi A_g \xi} \, \mathrm{e}^{\bar\xi f_-^\dagger + \xi f_-} \,,
\end{align}
which inherits the representation property $\sigma_-(gh) = \sigma_-(g)\, \sigma_-(h)$ from that of $\sigma_+\,$.

The bosonic versions of the formulas above have been described in
\cite{Howe1988} and \cite[Sect.~5]{Conrey2005}, and we refer to these
references for the details. Here we just report the result:
\begin{align}\label{eq:74mz}
  \omega_+(g) = \Det^{-1}(1-g) \int_{v_+} \me^{-\frac{1}{2} \bar{v}_+ A_g v_+} \, \me^{v_+ b_+^\dagger -  \bar{v}_+ b_+} \,,
\end{align}
where
\begin{align}
  \int_{v} = \pi^{-N} \int \prod_{\ell \in {\cal L}} \dd \mathrm{Re}\, v(\ell)\, \dd \mathrm{Im}\, v(\ell) .
\end{align}
We observe that
\begin{align}
    \mathrm{Re}\, A_g &= {\textstyle{\frac{1}{2}}} \big( A_g + A_g^\dagger \big) \\ &= {\textstyle{\frac{1}{2}}} (1-g^\dagger)^{-1} \left( (1-g^\dagger)(1+g) + (1+g^\dagger)(1-g) \right) (1-g)^{-1} \\
    &= (1-g^\dagger)^{-1} (1-g^\dagger g) (1-g)^{-1} .
\end{align}
Thus for $g$ a contraction ($1 - g^\dagger g > 0$) one has $\mathrm{Re}\, A_g > 0$, which makes the integral \eqref{eq:74mz} converge. The advanced representation is
\begin{align}
  \omega_-(g) = \Det^{-1}(g-1) \int_{v_-} \me^{+ \frac{1}{2}\bar{v}_- A_g v_-} \, \me^{\bar{v}_- b_-^\dagger - v_- b_-} ,
\end{align}
and this converges for $\mathrm{Re}\, A_g < 0$ or, equivalently, $g^\dagger g > 1$. By manipulations similar to those performed in the fermionic case one can prove the property $\omega_{\pm}(gh) = \omega_{\pm}(g)\, \omega_{\pm}(h)$.

\subsection{Second-quantized time-evolution operator}
\label{sec:2nd_quant}

Guided by our goal to derive formulas for the disorder-averaged observables of Sect.\ \ref{sect:obs}, we now put together the four representations (retarded/advanced for bosons/fermions). To keep the derivation straightforward and make the emerging supersymmetries manifest, we wish to perform manipulations such as $(1 - T^\dagger)^{-1} = - {T^{-1}}^\dagger (1 - {T^{-1}}^\dagger)^{-1}$. These are not immediately possible, however, as $Q$ and hence $T^\dagger = U^{-1} Q$ do not have inverses as operators on $\Hbulk \oplus \Hcontact$. We therefore regularize $Q$, replacing it by an operator $Q_\epsilon$ that does have the property of being invertible. We choose
\begin{align}\label{eq:regQ}
  Q_\epsilon = \me^{-\epsilon}Q + (1 - \me^{-\epsilon}) P ,
\end{align}
with small $\epsilon > 0$. Thus, for now, we take the network-model time evolution operators to be $T = Q_\epsilon U$ and $T^\dagger = U^{-1} Q_\epsilon\,$. We will eventually take the limit $\epsilon \to 0$.

Physically speaking, a small $\epsilon$ introduces weak absorption on all links in the bulk and kills only a reduced fraction $1-\me^{-\epsilon}$ of the flux arriving in the outgoing semi-links. Beyond making $T^\dagger$ invertible, the regularization (\ref{eq:regQ}) guarantees that
\begin{align}
    T T^\dagger = (Q_\epsilon\UU) (Q_\epsilon\UU)^\dagger = (Q_\epsilon)^2 = \me^{-2\epsilon}Q + (1-\me^{-\epsilon})^2 P < 1 ,
\end{align}
which ensures the convergence of the trace of $\omega_+ (T)$ for $T = Q_\epsilon \UU$. By the same token, the convergence of the trace of $\omega_-( {T^{-1}}^\dagger)$ for ${T^{-1}}^\dagger = Q_\epsilon^{-1} \UU$ is guaranteed.

At this point, it is useful to enlarge our framework and consider the
space
\begin{align}\label{eq:W}
  W = \mathcal{H} \otimes (\mathbb{C}^+\oplus \mathbb{C}^-)\otimes \mathbb{C}^{1|1}
  = W_{0}^{+}\oplus  W_{0}^{-}\oplus W_{1}^{+}\oplus  W_{1}^{-} .
\end{align}
Here the subscript $0,1$ refers to the even and odd subspaces of a
$\mathbb{Z}_2$-graded vector space (see e.g.\ the introduction of
\cite{Conrey2005} for a summary of useful concepts from supersymmetry that we are going to use here). The indices $\pm$ stand for retarded and advanced blocks. Note that each summand $W_{0,1}^{\pm}$ is isomorphic to ${\cal H}$.
We write $\mathcal{F} = \wedge(W_1^+ \oplus W_1^-) \otimes S(W_0^+ \oplus W_0^-)$ for the tensor product of the fermionic and bosonic Fock spaces made from these summands of $W$.

Operators on $W$ have a $4\times 4$ block structure by the decomposition above. We introduce the block-diagonal operators
\begin{align}
  \widehat{Q}_\epsilon &= \diag\left( Q_\epsilon , Q_\epsilon^{-1} ,
    Q_\epsilon , Q_\epsilon^{-1} \right) ,\\
  \widehat{\UU}_\epsilon &= \widehat{Q}_\epsilon\cdot (\UU \otimes \id_4) =
  \diag\left( Q_\epsilon \UU, Q_\epsilon^{-1} \UU, Q_\epsilon \UU, Q_\epsilon^{-1} \UU\right) . \label{eq:hatUeps}
\end{align}
The second-quantized representation of the time-evolution operator is then defined as the following product:
\begin{align}
  \label{eq:rho}
  \rho(\widehat{\UU}_\epsilon) &=
  \omega_+\left(Q_\epsilon \UU\right)\omega_-\left(Q_\epsilon^{-1} \UU \right)
  \sigma_+\left(Q_\epsilon \UU\right)\sigma_-\left(Q_\epsilon^{-1} \UU \right) \\
  &=
  \me^{b_+^\dagger \log(Q_\epsilon \UU) b_+}\,
  \me^{-b_- \log(Q_\epsilon^{-1} \UU) b_-^\dagger}\,
  \me^{f_+^\dagger \log(Q_\epsilon \UU) f_+}\,
  \me^{f_- \log(Q_\epsilon^{-1} \UU) f_-^\dagger} .
  \label{eq:rho2}
\end{align}
As mentioned above, the choice of regularization $Q_\epsilon\UU$, $Q_\epsilon^{-1} \UU$ ensures the convergence of the traces of $\omega_\pm$ over Fock space. In contrast, the choice of $Q_\epsilon\UU$, $Q_\epsilon^{-1} \UU$ in the fermionic part is not forced by convergence but by the requirement that for any $\epsilon > 0$ the operation of tracing $\rho( \widehat{\UU}_\epsilon)$ over the fermionic and bosonic Fock spaces gives unity:
\begin{align}\label{eq:SUSYtrace}
  \STr \rho(\widehat{\UU}_\epsilon) = 1 .
\end{align}
This property is at the heart of the supersymmetry method.

\subsection{Supersymmetry of the model}

Using the representation property of $\omega_\pm$ and $\sigma_\pm$ we now separate the second-quantized time-evolution operator of the open network into two factors:
\begin{align}
  \rho(\widehat{\UU}_\epsilon) = \rho(\widehat{Q}_\epsilon) \,
  \rho(\UU \otimes \id_4) .
\end{align}
Note that $\rho(\UU \otimes \id_4)$ is the second-quantized time-evolution operator of the closed network. This operator does exist, but it would not have a finite trace as an isolated operator. Finiteness of the trace is brought about by the other factor, $\rho(\widehat{Q}_\epsilon)$.

An easy computation shows that $\lim_{\epsilon\to 0} \rho( \widehat{Q}_\epsilon)$ yields the unit operator in the bulk and the projector on the Fock vacuum $| \mathrm{vac} \rangle_{c_i}$ for each contact link $c_i :$
\begin{align}\label{eq:QtoPi}
  \pi(\{c_i\})\equiv \lim_{\epsilon\to 0} \rho(\widehat{Q}_\epsilon) =
   \lim_{\epsilon\to 0} \epsilon^{ \sum_{i=1}^r (b_+^\dagger b_+^{\vphantom{\dagger}} + f_+^\dagger f_+^{\vphantom{\dagger}} + b_-^\dagger b_-^{\vphantom{\dagger}} + f_-^\dagger f_-^{\vphantom{\dagger}}) (c_i)} = \prod_{i=1}^r | \mathrm{vac} \rangle\langle \mathrm{vac} |_{c_i} \,.
\end{align}
Thus the point contacts translate to operator insertions under the trace over Fock space, where the inserted operators restrict the sum over Fock states on the contact links to a single state, namely the Fock vacuum. We will write $\pi(c) \equiv \pi(\{ c_i \})$ for short.

Let us add a word of caution here. It may happen for special disorder configurations (see the discussion at the end of Sect.\ \ref{sect:2.2}) that the trace-class property of $\rho(\widehat{Q}_\epsilon) \rho(\UU \otimes \id_4)$ is lost in the limit $\epsilon \to 0$. However, such events are expected to occur with probability zero in the ensemble of disorder configurations assumed. Therefore, the Fock trace should remain well-defined and finite when the limit $\epsilon \to 0$ is taken \emph{after} the disorder average (see Sect.\ \ref{sect:4.1}).

We now exhibit some important global symmetries of the second-quantized version of the network model. To do so, we unify our framework by introducing the map
\begin{align}
  &\dd \rho :\; \gl(W) \to  \gl({\cal F}) , \label{eq:rho*} \\
  &\dd \rho(X) = \sum_{\ell,\ell^\prime} \sum_{a,b = 0}^3 c^\ast_a(\ell) \langle \ell \vert X_{ab} \vert \ell^\prime \rangle c_b(\ell^\prime) \equiv c^\ast X c , \label{eq:rho*2}
\end{align}
where our unified notation for Fock operators is
\begin{align}
    &c_0^\ast = b^\dagger_+ , \quad
    c_1^\ast = f_+^\dagger , \quad
    c_2^\ast = - b_- , \quad
    c_3^\ast = f_- , \label{eq:104-mz} \\
    &c_0 = b_+ , \quad
    c_1 = f_+ , \quad
    c_2 = b_-^\dagger , \quad
    c_3 = f_-^\dagger . \label{eq:105-mz}
\end{align}
(Here we are using retarded-advanced ordering instead of boson-fermion ordering.)

Using the (Lie superbracket) commutation relations $[ c_a(\ell) , c_b^\ast(\ell^\prime)] = \delta_{ab} \delta_{\ell \ell^\prime}$ for the Fock operators $c$ and $c^\ast$ one can easily check that the mapping \eqref{eq:rho*} furnishes a Lie superalgebra representation of $\gl(W)$ on Fock space:
\begin{align}
  [\dd \rho(X) , \dd \rho(Y)] = \dd\rho([X,Y]) .
\end{align}
Note that $\dd\rho$, when restricted to block-diagonal operators, coincides with the infinitesimal version of $\rho$ as defined by \eqref{eq:rho}:
\begin{align}
  \dd\rho \big\vert_{W_0^\pm\to W_0^\pm} = \dd \omega_{\pm} \,, \quad
  \dd \rho \big\vert_{W_1^\pm\to W_1^\pm} = \dd \sigma_\pm \,.
\end{align}
In view of the tensor-product structure, it is now evident that for any constant element $X \in \gl_{2|2}$ the operator
\begin{align}
  \label{eq:rhostar}
  \dd\rho(\id_N \otimes X) = \sum_{\ell\in {\cal L}} \sum_{ab}
  c_a^\ast(\ell) X_{ab}\, c_b(\ell)
\end{align}
commutes with $\rho(\UU \otimes \id_4)$:
\begin{align}
  [\dd\rho (\id_N \otimes X) , \rho(\UU \otimes \id_4)] = 0 .
\end{align}
In this sense $\gl_{2|2}$ is a Lie superalgebra of global symmetries of the second-quantized formulation of the (closed) network model. Let us note that there exists a real Lie subalgebra $\mathfrak{u}_{1,1} \oplus \mathfrak{u}_2 \subset \gl_{2|2}$ which is represented (via $\dd\rho$) by anti-Hermitian operators and exponentiates to a Lie group $\mathrm{U}(1,1) \times \mathrm{U}(2)$ acting by unitary operators on the Fock space of the network model. Some degrees of freedom in $\mathrm{U}(1,1)$ are non-compact.

We also note that the point-contact operator $\pi$ breaks some of the $\gl_{2|2}$ symmetries including all of the non-compact subgroups of $\mathrm{U}(1,1)$ (this is crucial in order for the Fock trace of $\rho( \widehat{\UU}_\epsilon)$ to converge to a finite value in the limit $\epsilon \to 0$, after disorder averaging). In fact, $\pi$ is invariant only under the subalgebra $\fk \subset \gl_{2|2}$ of operators that are built (in the representation by $\dd \rho$) from one creation ($b_\pm^\dagger\,$, $f_\pm^\dagger$) and one annihilation operator ($b_\pm\,$, $f_\pm$). In the decomposition by retarded and advanced sectors (instead of boson and fermion sectors) this is the subalgebra
\begin{align}\label{eq:fk}
    \fk = \gl_{1|1} \oplus \gl_{1|1}
\end{align}
of block-diagonal matrices.


\section{The vertex model}\label{sect:4}

In this section we will show that upon performing the disorder average in the second-quantized formulation of the network model, one obtains a supersymmetric (SUSY) vertex model with infinite-dimensional $\mathfrak{gl}_{2|2} $-representations on the links.

\subsection{Disorder average}\label{sect:4.1}

Assuming the second-quantized formulation of the network model as described above, let now $O$ be a local operator (or a product of local operators), i.e.\ an expression built from the boson and/or fermion creation and annihilation operators of a small region (or several small and well separated regions) of the network. We will assume $O$ to be $\mathrm{U}(1)$-invariant in a sense to be explained below.

We recall that $\UU = \UUr \UUs\,$, and we write $\UUr \otimes \id_4 = \widehat{U}_\mathrm{r}\,$, $\UUs \otimes \id_4 = \widehat{U}_\mathrm{s}\,$, and $\UU \otimes \id_4 = \widehat{\UU}$. Before disorder averaging, the statistical average of $O$ is defined as
\begin{align}\label{eq:stat-av}
  \langle O \rangle_\mathcal{F} := \STr \pi(c)\rho(\widehat{\UU}) O ,
\end{align}
where $\pi(c) \rho(\widehat{\UU})$ is the limit ($\epsilon \to 0+$) of $\rho( \widehat{\UU}_\epsilon)$ in \eqref{eq:rho2}. Note that in view of Eq.\ \eqref{eq:SUSYtrace} this is a normalized statistical trace.

We now take the disorder average (denoted by $\mathbb{E}$) of the statistical trace above:
\begin{align}
  \mathbb{E}\, \STr \pi(c) \rho(\widehat{\UU}) O .
\end{align}
Recalling that $\rho$ is a group representation, we factorize $\rho( \widehat{\UU}) = \rho(\widehat{U}_\mathrm{r}) \rho(\widehat{U}_\mathrm{s} )$. The disorder average is then easily performed to give
\begin{align}
  \label{eq:EtoV}
  \mathbb{E}\, \STr \pi(c) \rho(\widehat{\UU}) O
  = \STr \mathcal{P} \rho(\widehat{U}_\mathrm{s}) O \pi(c) ,
\end{align}
where
\begin{align}
  \mathcal{P} \equiv \mathbb{E}\, \rho(\widehat{U}_\mathrm{r}) =
  \prod_{\ell} \int_0^{2\pi}\frac{\dd \phi_\ell}{2\pi}\,
  \me^{i\phi_\ell(n_+ - n_-)(\ell)}\, ,\quad
  n_\pm =  b_\pm^\dagger b_\pm^{\vphantom{\dagger}} +
    f_\pm^\dagger f_\pm^{\vphantom{\dagger}} \,,
\end{align}
projects the Fock space at each link $\ell$ to the subspace $V_\ell$ of states with an equal number of particles in the advanced and retarded sectors:
\begin{align}
  V_\ell = \mathrm{span}_\mathbb{C} \left\{ |0\rangle,  \,
  (b_+^\dagger b_-^\dagger) (\ell) |0\rangle,\,
  (b_+^\dagger f_-^\dagger) (\ell) |0\rangle,\,
  (f_+^\dagger b_-^\dagger) (\ell) |0\rangle,\,
  (f_+^\dagger f_-^\dagger) (\ell) |0\rangle,\, \dots \right\} .
\end{align}
Here $\vert 0 \rangle \equiv \vert \mathrm{vac} \rangle_\ell$ is the Fock vacuum at the link $\ell$. The assumed $\mathrm{U}(1)$-invariance of $O$ means that $O$ commutes with $\me^{\mathrm{i} \phi_\ell(n_+ - n_-) (\ell)}$ for all $\ell$ and hence with the projection operator $\mathcal{P}$. If we denote the total tensor product of spaces by ${\cal V}$,
\begin{align}
  {\cal V} = \bigotimes_\ell V_\ell \,,
\end{align}
we have
\begin{align}\label{eq:116-mz}
  \mathbb{E}\STr \pi(c) \rho(\widehat{\UU}) O
  = \STr_{\cal V} \rho(\widehat{U}_\mathrm{s}) O \pi(c) \equiv
  \langle O \rangle_\mathcal{V}\, .
\end{align}
The statistical sum on the right-hand side will turn out to be that of a vertex model.

\subsubsection{The representation $V$}

We now discuss the representation space $V \equiv V_\ell$ for a fixed link of the lattice, suppressing for the moment the link index $\ell$. $V$ is an irreducible lowest-weight representation of the Lie superalgebra $\gl_{2|2}\,$. To verify the lowest-weight property, notice that by the definition (\ref{eq:rho*2}) all operators $\dd \rho(X)$ that come from a strictly lower-triangular matrix $X$ annihilate the Fock vacuum. Thus if we take for the Cartan subalgebra of $\mathfrak{gl}_{2|2}$ the diagonal matrices and for the lowering operators the strictly lower-triangular matrices, then the Fock vacuum is a lowest-weight vector for the $\mathfrak{gl}_{2|2}$-representation $\dd \rho$ on $V$.

Moreover, all states of the representation $V$ are obtained by repeatedly acting with raising operators $\dd \rho(X)$ (from strictly upper-triangular matrices $X$) on the Fock vacuum. (Note that not all of the raising operators act non-trivially, since the Hilbert ray of $|0\rangle$ is invariant under the subalgebra $\fk$ of \eqref{eq:fk}). Conversely, from every state of $V$ one can reach the Fock vacuum by repeatedly acting with lowering operators. This means that the representation is irreducible. Since an arbitrary number of boson pairs can be created by multiple actions of $b_+^\dagger b_-^\dagger$, the representation space $V$ is infinite-dimensional.

\begin{figure}[h]
  \centering
  \begin{tikzpicture}[scale=1]
  \node at (0,0) {$V$ : };
  \begin{scope}[xshift=1.25cm,yshift=-.5]
    \node at (0.75,1.5) {{\scriptsize $n_f-1$}};
    \node at (6,-.25) {{\scriptsize $n_b+1$}};
    \draw[->] (-0.5,0)--(5.5,0);
    \draw[->] (0,-1.5)--(0,1.5);
    \foreach \x in {1,2,3,4,5} 
    { 
      \draw (\x,-0.1)--(\x,.1);
    }
    \draw (-0.1,-1)--(.1,-1);
    \draw (-0.1,1)--(.1,1);
    \filldraw (1,-1) circle (2pt) node[below] {{\scriptsize
        $|0\rangle$}};
    \foreach \x in {1,3} 
    { 
      \foreach \y in {1} 
      {
        \filldraw (\x,\y) circle (2pt);
        \filldraw (\x+2,\y-2) circle (2pt);
      } 
    }
    \foreach \x in {2,4} 
    { 
      \filldraw (\x-.1,0) circle (2pt);
      \filldraw (\x+.1,0) circle (2pt);
    }
    \begin{scope}[rotate=45,yshift=-1.4cm,xshift=-.6cm]
      \draw[] (2,0) ellipse (.5 and 1.75);
    \end{scope}
    \begin{scope}[xshift=2cm]
    \begin{scope}[rotate=45,yshift=-1.4cm,xshift=-.6cm]
      \draw[] (2,0) ellipse (.5 and 1.75);  
    \end{scope}
    \end{scope}
    \node at (2.5,-.5) {{\scriptsize  $W_1$}};
    \node at (2.5+2,-.5) {{\scriptsize $W_2$}};
  \end{scope}
  \node at (5+1.25,0.5) {$\dots$};
\end{tikzpicture}
  \caption{The weights of the representation $V$ in the two-dimensional lattice indexed by the number $n_f$ of fermions and $n_b$ of bosons. The four-dimensional subspaces $W_k$ ($k=1,2,\dots$) are $\fk$-multiplets.}
\label{fig:calv}
\end{figure}
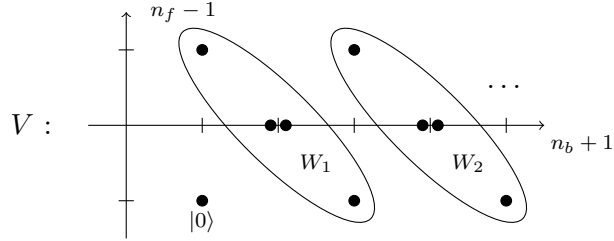

A simple graphical picture of $V$ is drawn in the two-dimensional lattice of Fig.\ \ref{fig:calv}. The axes are indexed by the boson number $n_b$ and fermion number $n_f\,$, each shifted by one unit. To motivate the shifts, note that $\mathfrak{gl}_{2|2}$ has two $\mathfrak{sl}_2$-subalgebras represented on $V$ by
\begin{align}
    &\mathfrak{sl}_2^\mathrm{bos} =
    \mathrm{span}_\mathbb{C} \left\{
    b_+^\dagger b_+^{\vphantom{\dagger}}
    + b_-^{\vphantom{\dagger}} b_-^\dagger \,,
    b_- b_+ \,, b_+^\dagger b_-^\dagger \right\} , \label{eq:sl2b}\\
    &\mathfrak{sl}_{2}^\mathrm{ferm} =
    \mathrm{span}_\mathbb{C} \left\{
    f_+^\dagger f_+^{\vphantom{\dagger}}
    - f_-^{\vphantom{\dagger}} f_-^\dagger \,,
    f_- f_+ \,, f_+^\dagger f_-^\dagger \right\} . \label{eq:sl2f}
\end{align}
The numbers $n_b + 1$ and $n_f - 1$ are the eigenvalues (or weights) of the Cartan subalgebra generators $b_+^\dagger b_+^{\vphantom{\dagger}} + b_-^{\vphantom{\dagger}} b_-^\dagger$ and $f_+^\dagger f_+^{\vphantom{\dagger}} - f_-^{\vphantom{\dagger}} f_-^\dagger$, respectively.

For good mathematical perspective, we here wish to mention the notion of a Howe pair \cite{cheng2012}. By definition, a Howe pair in our context consists of a Lie superalgebra $\mathfrak{g}$ and a classical Lie algebra $\mathfrak{h}$, with the property that $\mathfrak{g}$ and $\mathfrak{h}$ are mutual centralizers as subalgebras of an orthosymplectic Lie superalgebra. By the embedding into the latter, such a pair comes with a natural representation on a Fock space of bosons and fermions. The salient feature of a reductive Howe pair is that Fock space, viewed as a ($\mathfrak{g}, \mathfrak{h}$)-bimodule, has a multiplicity-free decomposition where each isomorphism class of $\mathfrak{g}$-irreducible representations that occur is associated with a unique isomorphism class of $\mathfrak{h}$-irreducible representations, and vice versa.

The Howe pair at hand consists of $\gl_{2|2}$ and $\gl_1$, where the latter coincides by accident with the center of $\gl_{2|2}$ (generated by the identity matrix in $\mathfrak{gl}_{2|2}$). The identity generator of $\gl_1$ is represented via $\dd \rho$ as the difference between the number of retarded-type and advanced-type particles, $n_+ - n_-\,$. By the stated Howe-pair property of multiplicity-freeness, our Fock space decomposes into sectors of fixed charge $n_+ - n_-$ for $\gl_1\,$, with each sector being acted upon irreducibly by $\gl_{2|2}\,$. The operation of taking the disorder average projects Fock space onto the $\gl_1$-singlet sector, which is $V$.

The two states with $n_b = 0$ constitute the fundamental representation of the $\mathfrak{sl}_2$ fermion-fermion subalgebra, while the states with fixed fermion number $n_f = 0$, $n_f = 1$ (retarded), $n_f = 1$ (advanced), and $n_f = 2$, form four discrete-series irreducible representations of the boson-boson subalgebra. All the states with even fermion number ($n_f \in \{0,2\}$) form a unitary irreducible representation of $\mathrm{U}(1,1) \times \mathrm{U}(2)$.

\subsubsection{The vertex structure}

We now move on to describe $\mathcal{P} \rho(\widehat{U}_\mathrm{s})$. We recall from Eq.\ \eqref{eq:UUsSv} that $\UUs$ is a direct sum of linear transformations $S_n$ ($n \in \mathcal{N}$), where $S_n$ relates incoming to outgoing states at the node $n$. Correspondingly, $\mathcal{P} \rho( \widehat{U}_\mathrm{s})$ can be written as
\begin{align}
  \mathcal{P} \rho(\widehat{U}_\mathrm{s}) = \bigotimes\nolimits_n {\cal R}_n \,, \quad {\cal R}_n :\; V_1 \otimes V_2 \to V_3 \otimes V_4 \,,
\end{align}
where $1,2$ (resp.\ $3,4$) label the two incoming (resp.\ outgoing) links at $n$. This formula tells us that the model which emerges after disorder averaging has the structure of what is called a vertex model.

To write an explicit expression for $\mathcal{R}_n\,$, we recall from Eqs.\ (\ref{eq:5-mz}, \ref{eq:6-mz}) that $S_n$ can be written as an endomorphism of $(\mathbb{C})_1 \oplus (\mathbb{C})_2$ with matrix
\begin{align}
    \begin{pmatrix} \mathrm{e}^{+\mathrm{i}\pi/4} \cos(t_n)
    &\mathrm{e}^{-\mathrm{i}\pi/4} \sin(t_n)\\
    \mathrm{e}^{-\mathrm{i}\pi/4} \sin(t_n)
    &\mathrm{e}^{+\mathrm{i}\pi/4} \cos(t_n) \end{pmatrix} = \mathrm{e}^{\mathrm{i}\pi/4} \, \mathrm{e}^{-\mathrm{i} t_n \sigma_1}
\end{align}
followed by the identity map $\mathrm{Id} : \; (\mathbb{C})_1 \oplus (\mathbb{C})_2 \to (\mathbb{C})_3 \oplus (\mathbb{C})_4$ with respect to the fixed choice of basis $\vert \ell \rangle \in (\mathbb{C})_\ell$ ($\ell = 1, \ldots, 4$). According to the machinery developed in Sect.\ \ref{ref:2nd}, the second-quantized form of the factor $\mathrm{e}^{-\mathrm{i} t_n \sigma_1}$ is
\begin{align} \label{eq:R_v}
  \rho(\id_4 \otimes \mathrm{e}^{-\mathrm{i} t_n \sigma_1} ) =
  \exp \left( -\mathrm{i} t_n \sum\nolimits_a \big( c_a^\ast (1) c_a^{\vphantom{\ast}} (2) +  c_a^\ast (2) c_a^{\vphantom{\ast}} (1) \big) \right) ,
\end{align}
where $c_a$ and $c_a^\ast$ ($a = 0, 1, 2, 3$) are the Fock operators defined in (\ref{eq:104-mz}, \ref{eq:105-mz}). Similarly one has
\begin{align}
  \rho(\id_4 \otimes \mathrm{e}^{\mathrm{i} \pi/4} \id_2 ) =
  \exp \left( (\mathrm{i} \pi/4) \sum\nolimits_a \big( c_a^\ast (1) c_a^{\vphantom{\ast}} (1) +  c_a^\ast (2) c_a^{\vphantom{\ast}} (2) \big) \right) .
\end{align}
Assembling the various factors we obtain
\begin{align}
    \mathcal{R}_n = \mathrm{Id}_{V_3 \otimes V_4 \leftarrow V_1 \otimes V_2} \times \mathcal{P}_{V_1 \otimes V_2} \times \rho(\id_4 \otimes \mathrm{e}^{-\mathrm{i} t_n \sigma_1} ) .
\end{align}
The term representing the scalar phase factor $\mathrm{e}^{\mathrm{i} \pi/4}$ has disappeared because the projection $\mathcal{P}$ trivializes $\sum_a c_a^\ast c_a^{\vphantom{\ast}} = n_+ - n_- = 0$.

The expression for $\mathcal{R}_n$ simplifies in the strongly anisotropic limit, where $t_n = \epsilon$ (resp.\ $t_n = \pi/2 - \epsilon$) on sublattice $A$ (resp.\ $B$) of the network (see Fig.\ \ref{fig:network}), and $\epsilon$ is taken to be very small. Expanding the exponential in \eqref{eq:R_v}, we see that the contribution of first order in $\epsilon$ vanishes due to the projection to $V_1 \otimes V_2\,$. Going up to second order one obtains
\begin{align}
  \mathcal{R}_n = \mathrm{Id}_{V_3 \otimes V_4 \leftarrow V_1 \otimes V_2} \left( 1 - 2 \epsilon^2 \sum\nolimits_{ab} c_a^\ast(1) c_b^{\vphantom{\ast}} (1) (-1)^{|b|} c_b^\ast (2) c_a^{\vphantom{\ast}} (2) + O(\epsilon^4) \right),
\end{align}
where $|b|$ stands for the $\mathbb{Z}_2$-degree of the index $b$, i.e.\ $|b| = 0$ for $b \in 2\mathbb{N}$ and $|b| = 1$ for $b \in 2\mathbb{N}+1$. The two-body operator (with coefficient $\epsilon^2$) is recognized as the $\gl_{2|2}$ quadratic Casimir operator realized on the tensor-product representation $V_1 \otimes V_2\,$.

\subsection{Highest-weight elements}
\label{ref:nilpotent}

We now introduce certain operators which will be related to observables of the network model in the next section. These operators are the vertex-model analogs of the sigma-model pure scaling fields discussed in \cite{Gruzberg2013}, and they appeared for the first time in our previous work \cite{BWZ}. We do not aim here at the derivation of the results of \cite{BWZ} in full generality, as this would take us too far from the concrete goals of the present paper. Instead, we will detail a subset of those results and briefly comment on the general case at the end. For now we fix a link and suppress the link label $\ell$.

To motivate the following construction, we begin with some Lie-theoretic perspective. Recall that the symmetry algebra $\gl_{2|2}$ contains two $\mathfrak{sl}_2$-subalgebras, one in the boson-boson and another one in the fermion-fermion sector. The main focus here will be on the former. For our present purposes, it is imperative to understand that the complex Lie algebra $\mathfrak{sl}_2^\mathrm{bos}$ of \eqref{eq:sl2b} contains a distinguished real form, $\mathfrak{su}_{1,1}\,$. This is the subalgebra of generators in $\mathfrak{sl}_2^\mathrm{bos}$ that are represented (via $\dd \rho$) by anti-Hermitian operators:
\begin{align}
    \mathfrak{su}_{1,1} = \mathrm{span}_\mathbb{R} \left\{
    \mathrm{i}( b_+^\dagger b_+^{\vphantom{\dagger}}
    + b_-^{\vphantom{\dagger}} b_-^\dagger) , \;
    \mathrm{i} (b_- b_+ + b_+^\dagger b_-^\dagger) , \;
    b_- b_+ - b_+^\dagger b_-^\dagger \right\} .
\end{align}
Its distinguishing feature is that the conjugation action on Fock operators by the associated non-compact Lie group $\mathrm{SU}(1,1)$ preserves not just the canonical commutation relations but also the relations under taking the Hermitian adjoint ($\dagger$) in Fock space.

Due to non-compactness, the Cartan subalgebras of $\mathfrak{su}_{1,1}$ organize into more than one conjugacy class \cite{Knapp2002}. In describing the representation space $V$, we made good use of the Cartan subalgebra generated by $\mathrm{i} ( b_+^\dagger b_+^{\vphantom{\dagger}} + b_-^{\vphantom{\dagger}} b_-^\dagger)$. The following discussion, however, has to be based on a different Cartan subalgebra, namely the one generated by
\begin{align}\label{eq:H-mz}
    H = \me^{-\mathrm{i}\alpha} \, b_+^\dagger b_-^\dagger - \me^{\mathrm{i}\alpha} \, b_- b_+
\end{align}
for some arbitrary (but fixed) angle $\alpha$. This Cartan subalgebra cannot be conjugate to the former one, as it lies in a different norm sector of the Cartan-Killing form for $\mathfrak{sl}_2\,$.

Adopting $H$ as our Cartan generator, we have the usual $\mathfrak{sl}_2$-commutation relations
\begin{align}\label{eq:sl2-mz}
    [H , E] = 2 E , \quad [H , F] = - 2 F , \quad [E , F] = H ,
\end{align}
for an associated pair $E$, $F$ of root vectors. We make the choice
\begin{align}
    E = B_+ C_- \,, \quad F = C_+ B_- \,,
\end{align}
where
\begin{align} \label{eq:BC}
    &B_+ = \frac{1}{\sqrt{2}} \left( b_+^\dagger - \me^{\mathrm{i} \alpha} b_- \right), \quad
    B_- = \frac{1}{\sqrt{2}} \left( b_+ + \me^{-\mathrm{i} \alpha}b_-^\dagger \right) , \\
    &C_+ = \frac{1}{\sqrt{2}} \left( b_+^\dagger + \me^{\mathrm{i} \alpha} b_- \right), \quad
    C_- = \frac{1}{\sqrt{2}} \left( b_+ - \me^{-\mathrm{i} \alpha}b_-^\dagger \right) .
\end{align}
(In \cite{BWZ} the operators $B_+, C_-$ were denoted by $B^\dagger , B$.) To verify the $\mathfrak{sl}_2$-commutation relations note that
\begin{align}
    [B_- , B_+] = 1, \quad [C_- , C_+] = 1, \quad H = B_+ B_- - C_+ C_- \,,
\end{align}
and $[B_\pm , C_\pm] = 0 = [B_\pm , C_\mp]$. Note also that $\mathfrak{su}_{1,1}$ is spanned by $H$, $\mathrm{i}E$, $\mathrm{i}F$.

The key property of the present root-space decomposition $\mathfrak{su}_{1,1} = \mathbb{R} \cdot H \oplus \mathbb{R} \cdot \mathrm{i}E \oplus \mathbb{R} \cdot \mathrm{i}F$ is that the root vector $E$ is positive (and so is, in fact, $F$). Indeed, we have
\begin{align}
    C_- = B_+^\dagger \,, \quad C_+ = B_-^\dagger ,
\end{align}
and hence $E = B_+ C_- = C_- B_+ = B_+^\dagger B_+^{\vphantom{\dagger}}$ is self-adjoint and non-negative. It is not difficult to see that there exists no vector in the $\gl_{2|2}$-representation space $V$ which is annihilated by $B_+$ or its adjoint. Hence we have strict positivity: $E > 0$. (Please note that no such property can be had with the ``wrong'' choice of Cartan generator as the boson number operator.) This feature will allow us to define $E^q$ for complex values of $q$.

Turning to the fermion-fermion sector, we note that in this case all Cartan subalgebras of the compact real form $\mathfrak{su}_2 \subset \mathfrak{sl}_2^\mathrm{ferm}$ are conjugate to one another and no serious harm would result from continuing to work with the Cartan subalgebra generated by the fermion number operator $\mathrm{i}( f_+^\dagger f_+^{\vphantom{\dagger}} - f_-^{\vphantom{\dagger}} f_-^\dagger) \in \mathfrak{su}_2\,$. However, it is desirable and useful for calculational purposes to preserve the symmetry between bosons and fermions. We therefore mimic the construction of the boson-boson sector.

With this motivation, we proceed to define fermionic counterparts for $B_\pm\,$, $C_\pm :$
\begin{align}
  F_+ = \frac{1}{\sqrt{2}}\left(f_+^\dagger + \me^{\mathrm{i} \alpha} f_- \right) , \quad
  F_- = \frac{1}{\sqrt{2}}\left(f_+ + \me^{-\mathrm{i} \alpha} f_-^\dagger \right) , \\
  G_+ = \frac{1}{\sqrt{2}}\left(f_+^\dagger - \me^{\mathrm{i} \alpha} f_- \right) , \quad
  G_- = \frac{1}{\sqrt{2}}\left(f_+ - \me^{-\mathrm{i} \alpha} f_-^\dagger \right) .
\end{align}
Their bracket relations are
\begin{align}
    [F_- , F_+] = 1, \quad [G_- , G_+] = 1,
\end{align}
and $[F_\pm , G_\pm] = 0 = [F_\pm , G_\mp]$. (Of course, in the present case of fermions the brackets are understood to mean the anti-commutator.) The relations under taking the Hermitian adjoint are now diagonal:
\begin{align}
    F_+ = F_-^\dagger \,, \quad G_+ = G_-^\dagger \,.
\end{align}
Altogether, the 16 operators made by multiplying one of the set $\{ B_+, C_+, F_+, G_+ \}$ with one of $\{ B_-, C_-, F_-, G_-\} $ form another basis for the Fock-space representation of $\mathfrak{gl}_{2|2}\,$. Truth be told, its one and only advantage compared with the original basis (\ref{eq:104-mz}, \ref{eq:105-mz}) is the presence of the strictly positive operator $B_+ C_-\,$.

In the remainder of this subsection, we focus on the quadratic expressions built by multiplying one of $\{B_+ , F_+\}$ with one of $\{ C_- , G_- \}$. By construction, these four combinations have vanishing Lie superbrackets (commutators or anti-commutators, as the case may be) with one another, and they make part of a nilpotent subalgebra of raising operators for the Lie superalgebra $\mathfrak{gl}_{2|2} \,$. Our goal now is to use this algebraic structure to construct \emph{highest-weight} elements; by which we mean elements in the universal enveloping algebra of $\mathfrak{gl}_{2|2}$ with the property that they are annihilated by the commutator action of every one from a set of raising operators. This is achieved as follows.

We arrange our plus/minus basis of Fock operators in a specific order,
\begin{align}
    \{ B_+ , F_+ , G_+ , C_+ \}
    \quad \text{and} \quad
    \{ B_- , F_- , G_- , C_- \} ,
\end{align}
and we take the four-dimensional Cartan subalgebra of $\mathfrak{gl}_{2|2}$ to be the algebra generated by
\begin{align}\label{eq:CA-mz}
    \{ B_+ B_- \,, \; F_+ F_- \,, \; G_+ G_- \,, \; C_+ C_- \} .
\end{align}
For the set of $\mathfrak{gl}_{2|2}$-raising operators we take the quadratic operators whose plus factor occurs earlier in the ordered sequence than the minus factor. These are the combinations
\begin{align}\label{eq:raising}
    B_+ F_- \,,\; B_+ G_- \,,\; B_+ C_- \,,\; F_+ G_- \,,\; F_+ C_- \,,\; G_+ C_- \,.
\end{align}
The remaining $4^2 - 4 - 6 = 6$ quadratic combinations (where the plus factor occurs later than the minus factor) constitute the set of lowering operators.

It is useful to think of all 16 operators as being arranged in a square matrix, with the four Cartan generators placed on the diagonal according to the ordering of (\ref{eq:CA-mz}), the six raising operators placed above the diagonal (in the arrangement prescribed by the same ordering scheme), and the six lowering operators below the diagonal (also ordered by the same scheme). The usefulness of this picture stems from the fact that the operation of replacing each operator by the corresponding matrix with entry $1$ in the position of the operator and zeroes everywhere else, is an isomorphism of Lie superalgebras, i.e.\ the Lie superbracket (or commutation) relations are preserved.

Now $B_+ C_-$ resides in the right upper corner of the $4 \times 4$ array. Therefore, in view of the isomorphism above, it is immediately clear that this operator has vanishing commutator with every one of the raising operators (\ref{eq:raising}). It is also an eigenvector of the commutator action of any linear combination of the Cartan generators. Thus it is what is called a highest-weight vector \cite{Knapp2002} in the adjoint representation of $\mathfrak{gl}_{2|2}\,$.

We now put in place the network-model structure, which has been ignored by our discussion so far. Following the blueprint of Sect.\ \ref{sect:obs}, we choose a positive measure $\ell \mapsto w(\ell)$ supported on a small observation region $\mathcal{R}$. We then consider the coarse-grained operator
\begin{align}\label{eq:Q-BC}
    &M_{BC} = \sum\nolimits_\ell w(\ell) \, (B_+ C_-)(\ell) .
    \end{align}
Clearly, this still is a highest-weight vector for the adjoint action of the $\mathfrak{gl}_{2|2}$ generators summed (with unit coefficients) over the observation region $\mathcal{R}$.

Next, for algebraic completeness, we seek a second operator which shares the highest-weight properties of $M_{BC}\,$. Given our fixed choice of root-space decomposition and raising operators, it is not possible to find another such element in the Lie superalgebra $\mathfrak{gl}_{2|2}$ itself; for that, we need to pass to its universal enveloping algebra. Hence, consider the combination
\begin{align}\label{eq:zero-hw}
    (B_+ C_-) (F_+ G_-) - (F_+ C_-) (B_+ G_-) .
\end{align}
On a single link (or single representation space $V$) this combination vanishes, since $B_+$, $F_+$, $C_-$, $G_-$ generate a graded-commutative algebra in which the expression \eqref{eq:zero-hw} is a difference of identical terms. However, it becomes non-zero when the network-model operation of coarse graining is brought into play. Let
\begin{align}
    M_{FG} = \sum\nolimits_\ell w(\ell) \, (F_+ G_-)(\ell)
\end{align}
be the coarse-grained version of $F_+ G_-$ (one could also take a coarse-graining measure different from that for $B_+ C_-$), and define $M_{BG}$, $M_{FC}$ in the same way by coarse graining the operators $B_+ G_-$, $F_+ C_-$, respectively. We then ask whether the expression
\begin{align}\label{eq:second-hw}
    M_{BC} \, M_{FG} - M_{FC} \, M_{BG}
\end{align}
has the desired highest-weight property. To begin with, it is obvious that \eqref{eq:second-hw} commutes with the four $\mathcal{R}$-summed $\mathfrak{gl}_{2|2}$-raising operators made from the graded-commutative algebra of $B_+, F_+, C_-, G_-$. It is also clear that \eqref{eq:second-hw} is an eigenvector of the commutator action by the $\mathcal{R}$-summed Cartan generators. What remains to be checked is that \eqref{eq:second-hw} commutes with the two raising operators $\sum_{\ell \in \mathcal{R}} B_+(\ell) F_-(\ell)$ and $\sum_{\ell \in \mathcal{R}} G_+(\ell) C_-(\ell)$. This can easily be verified by working out the commutators explicitly. For a more conceptual argument, one observes that the expression \eqref{eq:second-hw} is related to a superdeterminant:
\begin{align}
    M_{BC} \, M_{FG} - M_{FC} \, M_{BG} &=
    M_{BC}^2 \, \frac{M_{FG} - M_{FC} \, M_{BC}^{-1} \, M_{BG}}{M_{BC}}
    \nonumber \\ &=  M_{BC}^2 \, \mathrm{SDet}^{-1} \begin{pmatrix}
    M_{BC} &M_{BG} \\ M_{FC} &M_{FG} \end{pmatrix} . \label{eq:144-mz}
\end{align}
(Note that the superdeterminant is well-defined, as the algebra of $M_{BC}, M_{FG}, M_{BG}, M_{FC}$ is graded-commutative; in particular, one has $M_{BG} \, M_{FC} + M_{FC} \, M_{BG} = 0$ and $M_{FC}^2 = M_{BG}^2 = 0$. Note also that our strictly positive operator $M_{BC}$ does have an inverse.) Using some Grassmann variable $\xi$ to form the 1-parameter group $\exp\big( \xi \sum_\ell B_+(\ell) F_-(\ell) \big)$, the desired result follows by the isomorphism to $4 \times 4$ (super-)matrices and the multiplicativity of the superdeterminant. This completes our conceptual verification that \eqref{eq:second-hw} indeed is a highest-weight element.

Since the operation of taking the (graded) commutator satisfies the (graded) Leibniz product rule, products of powers of highest-weight elements still have the highest-weight property. Thus the operators (\ref{eq:Q-BC}) and  (\ref{eq:second-hw}) are elementary building blocks for the construction of more general highest-weight elements, say the operator
\begin{align}\label{eq:hwv2}
    \Phi_{q,\,p}(\mathcal{R}) = M_{BC}^q \left( M_{FG} - M_{FC} \, M_{BC}^{-1} \, M_{BG} \right)^p \,,
\end{align}
where $p$ has to be a non-negative integer while $q$ may be complex (by the positivity of $M_{BC}$ and fractional calculus or analytic continuation). The eigenvalue (or weight) $\lambda_{q,\,p}$ of $\Phi_{q,\,p}$ with respect to the system of Cartan generators (\ref{eq:CA-mz}) is seen to be
\begin{align}
  \lambda_{q,\,p} = (q , p, -p, -q) .
\end{align}
Moreover, the product of two highest-weight elements,
\begin{align}
   \Phi_{q_1,\,p_1}(\mathcal{R}_1) \, \Phi_{q_2,\,p_2}(\mathcal{R}_2) ,
\end{align}
is still a highest-weight element, with weight given by the sum $\lambda_{q_1 + q_2,\, p_1 + p_2}$ of the individual weights (here we assume that the unit-coefficient sum of $\mathfrak{gl}_{2|2}$-generators is extended so as to cover both observation regions $\mathcal{R}_1$ and $\mathcal{R}_2$). It is this ``abelian fusion'' that makes the correlation functions of highest-weight elements more tractable than those of other operators.

Finally, let us comment briefly on the more general situation depicted in Sect.\ \ref{sect:obs}. What we have described here will enable us (in Sect.\ \ref{sect:5}) to transcribe to the vertex-model setting the network-model observables $h_q(K)\, e_p(K)$. The more general observables of Sect.\ \ref{sect:obs} can be treated in a similar way by introducing replicas; cf.\ \cite{Gruzberg2013}.


\section{Relation between network and vertex-model observables}\label{sect:5}

We will now build on the previous sections to introduce certain correlation functions in the vertex model and establish their relation to observables in the network model. As a last preparation, we will introduce one more piece of mathematics.

\subsection{More on the representation $\rho$}\label{sect:5.1}

We present here the extension of the character formulas (\ref{eq:STrDet+}, \ref{eq:STrDet-}) and (\ref{eq:48mz}, \ref{eq:49mz}) to the more general case of operators mixing the different sectors (i.e., retarded and advanced bosons and fermions). First of all, the extension to the case of retarded and advanced fermions is immediate. In that case, one has from (\ref{eq:STrDet+}) the easy result
\begin{align}
    &\STr \, \exp \sum_{\ell\ell^\prime}
    \begin{pmatrix} f_+^\dagger(\ell) & f_-(\ell) \end{pmatrix}
    \begin{pmatrix} \langle \ell | X_{++} | \ell^\prime \rangle
    & \langle \ell | X_{+-} | \ell^\prime \rangle \\
    \langle \ell | X_{-+} | \ell^\prime \rangle
    & \langle \ell | X_{--} | \ell^\prime \rangle \end{pmatrix}
    \begin{pmatrix} f_+(\ell^\prime) \\ f_-^\dagger(\ell^\prime) \end{pmatrix} \\
    &\equiv \STr \, \exp
    \begin{pmatrix} f_+^\dagger & f_- \end{pmatrix}
    \begin{pmatrix} X_{++} &X_{+-} \\ X_{-+} &X_{--} \end{pmatrix}
    \begin{pmatrix} f_+ \\ f_-^\dagger \end{pmatrix}
    = (-1)^N \Det(1 - \me^X) , \label{eq:149-mz}
\end{align}
by making a particle-hole transformation (induced by a complex-linear transformation of Fock space) in the advanced sector: $f_- \leftrightarrow f_-^\dagger$, and then using that the supertrace $\STr (\bullet) = \Tr\, (-1)^{\sum f_+^\dagger f_+^{\vphantom{\dagger}} + \sum f_-^\dagger f_-^{\vphantom{\dagger}}} (\bullet)$ reacts with a change of sign factor $(-1)^N$ due to the reversal of fermion parity for odd $N$.

There exists a bosonic counterpart to this formula, which looks very similar:
\begin{align}
  \label{eq:Tr_Det_general}
  \Tr \exp \begin{pmatrix} b_+^\dagger & -b_- \end{pmatrix}
  \begin{pmatrix} X_{++} &X_{+-} \\ X_{-+} &X_{--} \end{pmatrix}
  \begin{pmatrix} b_+\\ b_-^\dagger \end{pmatrix}
  &= (-1)^N \Det^{-1}(1-\me^X) .
\end{align}
The proof, however, is more difficult, as there exists no operator on the bosonic Fock space that would induce a particle-hole transformation $b_- \leftrightarrow b_-^\dagger$. As a matter of fact, one must impose a restriction on $X$ in order for the Fock trace in (\ref{eq:Tr_Det_general}) to exist and the formula to hold true. The restriction is best formulated in terms of the group element $g = \me^X$ and reads (cf.\ \cite{Conrey2005})
\begin{align}\label{eq:contract}
    g^\dagger s\, g < s ,
\end{align}
where $s$ is the identity (resp.\ minus the identity) in the retarded (resp.\ advanced) sector. In the special case of $X_{+-} = X_{-+} = 0$ this condition reduces to $\mathrm{Re}\, X_{++} < 0 < \mathrm{Re}\, X_{--}\,$.

We now combine these formulas and give the extension to the super-setting.  First, we extend by linearity the representation $\dd \rho$ of Eq.\ \eqref{eq:rho*} to \emph{super}matrices, i.e.\ matrices whose entries in the odd blocks are Grassmann variables. We denote by $\widetilde{\mathfrak{g}}$ the Lie algebra obtained by tensoring the odd part of $\gl_N \otimes \gl_{2|2} = \gl_{2N|2N}$ with a Grassmann algebra generated by $2 N^2$ variables. Then for a supermatrix $\widehat{X} \in \widetilde{\mathfrak{g}}$ [with a boson-boson block that satisfies the restriction of (\ref{eq:contract})] the general character formula is:
\begin{align} \label{eq:STrSDet}
  \STr \me^{\dd\rho(\widehat{X})} = \SDet^{-1}\big(1- \me^{\widehat{X}} \big).
\end{align}
Given (\ref{eq:149-mz}) and (\ref{eq:Tr_Det_general}), its proof is just a simple exercise in manipulating power series that terminate after finitely many terms.

For completeness, we recall \cite{Berezin} that the (reciprocal of the) superdeterminant of a supermatrix with even blocks $A, D$ and odd blocks $B, C$ is defined as
\begin{align}
  \SDet^{-1} \begin{pmatrix} A &B \\ C &D \end{pmatrix}
  = \frac{\Det(D-CA^{-1}B)}{\Det(A)} .
\end{align}
This formula, which was already assumed in Eq.\ (\ref{eq:144-mz}), holds whenever $A$ is invertible. Let us also mention in passing that it is straightforward to extend the fermionic and bosonic Gaussian integral
representations of Sect.\ \ref{sec:gaussian} to the super-setting.

\subsection{Generating function of observables}\label{sect:5.2}

We now continue the line of development of Sect.\ \ref{ref:nilpotent}. Our goal is to relate the statistical average of the vertex-model operators $\Phi_{q, \, p}$ to the disorder average of the corresponding observables in the network model. This is conveniently done by employing a generating function on both sides of the correspondence, as follows.

Starting on the vertex-model side, let there be $n$ (for $n$ ``observation regions'') $2 \times 2$ supermatrices $Y_j\,$, each consisting of two ordinary parameters $Y_j^{00}\,, Y_j^{11}$ and two Grassmann variables $Y_j^{01}\,, Y_j^{10}\,$, and consider the operator
\begin{align}
    M(Y) = \sum_{j=1}^n \sum_{\ell \in \mathcal{R}_j} w_j(\ell)
    \begin{pmatrix} B_+(\ell) & F_+(\ell) \end{pmatrix}
    \begin{pmatrix} Y_j^{00} &Y_j^{01}\\ Y_j^{10} &Y_j^{11} \end{pmatrix}
    \begin{pmatrix} C_- (\ell) \\ G_-(\ell) \end{pmatrix} ,
\end{align}
with weights $w_j(\ell) > 0$. The symbol ${\cal R}_j$ stands for the set of links in one of $n$ small observation regions of the network. We adopt the standard convention that the Grassmann variables $Y_j^{01}\,, Y_j^{10}$ anti-commute with the fermionic operators $F_+(\ell)$ and $G_-(\ell)$.

Our generating function is the vertex-model trace of the exponential of $M(Y)$:
\begin{align}\label{eq:genfunc}
    Z(Y) := \big\langle \me^{M(Y)} \big\rangle_\mathcal{V} =
    \STr_{\cal V} \pi(c) \rho(\widehat{U}_\mathrm{s})\, \me^{M(Y)} .
\end{align}
To convert this into a network-model quantity, we first of all re-introduce the disorder average $\mathbb{E}$, returning in Eq.\ \eqref{eq:116-mz} to the left-hand side:
\begin{align}
  Z(Y) = \mathbb{E}\, \STr \pi(c) \rho(\widehat{\UU}) \, \me^{M(Y)} =
  \lim_{\epsilon\to 0+} \mathbb{E}\, \STr \rho(\widehat{U}_\epsilon) \, \me^{M(Y)} .
\end{align}
The second equality trades the point-contact operator $\pi(c)$ for our regularization procedure with parameter $\epsilon = 0+$; cf.\ Eqs.\ \eqref{eq:hatUeps} and \eqref{eq:rho}.

The plan now is to use the character formula \eqref{eq:STrSDet}. For that purpose, we observe that $M(Y)$ can be written as the second quantization \eqref{eq:rho*2} of a supermatrix-valued operator $\widehat{Y}_w \in \widetilde{\mathfrak{g}}$:
\begin{align}
    &M(Y) = \dd\rho(\widehat{Y}_w), \quad \widehat{Y}_w = \sum w_j \otimes (E_\alpha / 2) \otimes Y_j \,, \\
    &w_j = \sum \vert \ell \rangle \, w_j(\ell) \, \langle \ell \vert , \quad
    E_\alpha = \begin{pmatrix} 1 \\ \me^{\mathrm{i} \alpha} \end{pmatrix}
    \otimes \begin{pmatrix} 1 &-\me^{-\mathrm{i} \alpha} \end{pmatrix} =
    \begin{pmatrix} 1 &-\me^{-\mathrm{i}\alpha} \\ \me^{\mathrm{i} \alpha} &-1 \end{pmatrix} .
\end{align}
In specifying $\widehat{Y}_w$ we have used the isomorphism $\mathfrak{gl}_{2N|2N} = \mathfrak{gl}_N \otimes \mathfrak{gl}_2 \otimes \mathfrak{gl}_{1|1}$ which factors the retarded-advanced structure ($\mathfrak{gl}_2$) out of the super-structure ($\mathfrak{gl}_{1|1}$) and the network model structure ($\mathfrak{gl}_N$). The matrix $E_\alpha / 2$ acting in the retarded-advanced sector appears because \eqref{eq:STrSDet} requires that the Fock operators $B_+ , F_+, C_-, G_-$ be expressed in terms of the original basis (\ref{eq:104-mz}, \ref{eq:105-mz}) assumed in \eqref{eq:rho*2}. Note that $E_\alpha$ is nilpotent: $E_\alpha^2 = 0$.

We now use the representation property $\rho(\widehat{U}_\epsilon) \, \me^{ \dd\rho(\widehat{Y}_w)} = \rho(\widehat{U}_\epsilon \, \me^{ \widehat{Y}_w})$ and apply the character formula \eqref{eq:STrSDet} to obtain
\begin{align}\label{eq:160-mz}
  Z(Y) = \lim_{\epsilon\to 0+} \mathbb{E}\, \STr \rho(\widehat{U}_\epsilon) \, \me^{\dd\rho(\widehat{Y}_w)} = \lim_{\epsilon\to 0+} \mathbb{E}\, \SDet^{-1} \big(1 - \widehat{U}_\epsilon \, \me^{\widehat{Y}_w} \big) .
\end{align}
Here we recall that $\STr$ means the supertrace over the tensor product of four Fock spaces (retarded and advanced, bosonic and fermionic), whereas $\SDet$ is a superdeterminant on the first-quantized Hilbert space $W = \mathcal{H} \otimes (\mathbb{C}^+ \oplus \mathbb{C}^-) \otimes \mathbb{C}^{1|1}\,$.

To evaluate the expression \eqref{eq:160-mz} further, we use two elementary facts. The first one is the nilpotency of $E_\alpha\,$, which entails $\widehat{Y}_w ^2 = 0$. The second one is that
\begin{align}
  \SDet^{-1} \big( 1 - \widehat{U}_\epsilon \big) = 1
\end{align}
due to the definition $\widehat{U}_\epsilon= \diag( Q_\epsilon \UU, Q_\epsilon^{-1} \UU) \otimes \mathbf{1}_{1|1}\,$. We thus obtain
\begin{align}\label{eq:162-mz}
  Z(Y) = \lim_{\epsilon\to 0+} \mathbb{E}\, \SDet^{-1} \big( 1 - (1-\widehat{U}_\epsilon)^{-1} \widehat{U}_\epsilon \widehat{Y}_w \big) .
\end{align}
Next, recalling the definition $T = Q \UU$, we carry out the limit $\epsilon \to 0+$:
\begin{align}\label{eq:164-mz}
    \lim_{\epsilon \to 0+} (1-\widehat{U}_\epsilon)^{-1} \widehat{U}_\epsilon
    = \diag\left( (1-T)^{-1} T , -(1-T^\dagger)^{-1} \right) \otimes \mathbf{1}_{1|1} \,.
\end{align}

Since the nilpotent matrix $E_\alpha$ in the retarded-advanced sector has rank one, the superdeterminant in \eqref{eq:162-mz} can be reduced further. A quick (if formal) way to arrive at the reduced expression is to use $\SDet = \exp \circ \STr \circ \log$ along with the power series of the logarithm $x \mapsto \log(1-x)$ and the identity
\begin{align}
    \mathrm{Tr}_{\mathbb{C}^+ \oplus \mathbb{C}^-} ( E_\alpha D)^k = \big( \mathrm{Tr}_{\mathbb{C}^+} D - \mathrm{Tr}_{\mathbb{C}^-} D \big)^k
\end{align}
for any operator $D$ that is diagonal in the retarded-advanced space $\mathbb{C}^+ \oplus \mathbb{C}^-$. Applying this to $(1 - \widehat{U}_\epsilon)^{-1} \widehat{U}_\epsilon \widehat{Y}_w$ in the limit given by \eqref{eq:164-mz}, we obtain the formula
\begin{align}
    Z(Y) = \mathbb{E}\, \SDet^{-1} \left( 1 - {\textstyle{\frac{1}{2}}} \sum\nolimits_j (G\cdot w_j) \otimes Y_j \right) ,
\end{align}
where $G$ is the network-model Green operator introduced in Eq.\ \eqref{eq:Green-op} and expressed in terms of $T$ in \eqref{eq:MEofK}. Note that the superdeterminant is now taken over the reduced space $\mathcal{H} \otimes \mathbb{C}^{1|1}\,$. The final step is convert the determinant over $\mathcal{H} \otimes \mathbb{C}^{1|1}$ to a determinant over $\mathbb{C}^r \otimes \mathbb{C}^{1|1}\,$, by passing from $G \cdot w$ to the $r \times r$ matrices $K(x_j)$ of Eq.\ \eqref{eq:30-mz}:
\begin{align}\label{eq:master}
    Z(Y) = \mathbb{E}\, \SDet^{-1} \left( 1 - {\textstyle{\frac{1}{2}}} \sum\nolimits_j K(x_j) \otimes Y_j \right) .
\end{align}
In summary, the formula \eqref{eq:master} re-expresses the vertex-model generating function \eqref{eq:genfunc} with parameters $Y$ as a generating function for multi-point correlation functions of the local wavefunction data $K(x_1), \ldots, K(x_n)$ in the network model. It is a key result of the present work.

\subsection{Correlation functions}\label{sect:5.3}

We will now specialize the equality between the generating functions \eqref{eq:genfunc} and \eqref{eq:master} to express certain correlators in the vertex model as network-model observables. The continuum limit of these correlation functions will be analyzed in Sect.\ \ref{sec:6}.

As a disclaimer, we should mention that the network-model observables to be obtained in the present section constitute only a subset of those introduced in Sect.\ \ref{sect:obs}. To handle the whole class of observables of Sect.\ \ref{sect:obs}, one would need to replicate the basic Fock operators for bosons and fermions. Although replica indices are easily incorporated into the present formalism, we will continue to concentrate on the minimal theory which we have developed so far.

To begin the discussion of correlation functions, a first remark is that since the theory \emph{before} taking the disorder average is a theory of free bosons and fermions, correlation functions obey Wick's theorem. One has
\begin{align}
  \begin{split}
      \Big< (B_+ C_-)(\ell_1) \cdots (B_+ C_-)(\ell_k) (F_ + G_-) (\ell_1') \cdots (F_ + G_-)(\ell_m') \Big>_{{\cal V}} = \\
      \mathbb{E}\; \text{Perm} \Big( \big< B_+ (\ell_i) C_-(\ell_j) \big>_{\mathcal{F}} \Big)_{i,j=1}^k \Det \Big( \big< F_+(\ell_i') G_-(\ell_j') \big>_{\mathcal{F}} \Big)_{i,j = 1}^m ,
  \end{split}
\end{align}
with Perm the permanent, and the two-point functions given by
\begin{align}
  \label{eq:two_point}
  \left< B_+(\ell) C_- (\ell^\prime) \right>_{\cal F} = - \left< F_+ (\ell) G_-(\ell^\prime) \right>_{\cal F} = \frac{1}{2} \sum_{i=1}^r \psi_{c_i}(\ell) \bar\psi_{c_i} (\ell^\prime) .
\end{align}
At this point, we recall the definitions \eqref{eq:stat-av} and \eqref{eq:116-mz} of the Fock-statistical averages before and after taking the disorder average. However, while Wick's theorem in principle allows one to compute any correlation function, it is simpler in practice to start directly from Eq.\ \eqref{eq:master}, which conveniently takes care of the combinatorics needed to resum all Wick contractions.

In the next two sections, we will discuss explicitly the case of the bosonic and fermionic operators introduced in Sect.\ \ref{ref:nilpotent}. Correlators of the general highest-weight operator $\Phi_{q,\,p}$ in \eqref{eq:hwv2} can be treated in a similar way.

\subsubsection{Bosonic observables}
\label{sec:bos_obs}

Consider the following correlation function of bosonic highest-weight elements:
\begin{align}
  f_{q_1,\dots,\,q_n}(c\,;x_1,\dots,x_n) &= \left<
  M_{BC}(x_1)^{q_1} \cdots M_{BC}(x_n)^{q_n} \right>_{\cal V} \label{eq:171-mz} \\
  &= \STr_{\cal V} \pi(c) \rho(\widehat{U}_\mathrm{s})\,
  M_{BC}(x_1)^{q_1} \cdots M_{BC}(x_n)^{q_n} ,
\end{align}
where, as we recall from Sect.\ \ref{ref:nilpotent}, $M_{BC}$ is a sum of strictly positive operators,
\begin{align}
     M_{BC}(x_j) = \sum_{\ell\in {\cal R}_j} w_j(\ell)\, B_+(\ell) C_-(\ell) ,
\end{align}
with weights $w_j(\ell) > 0$. As before, the symbol ${\cal R}_j$ stands for the set of links in a small observation region of the network. The position argument $x_j$ of $M_{BC}(x_j)$ is to indicate the coarse-grained location of the small region ${\cal R}_j$ in the continuum limit. Note, however, that all formulas here and in the following sections hold exactly (without assuming any continuum approximation).

The type of correlator \eqref{eq:171-mz} can be extracted from the generating function $Z(Y)$ for $Y_j^{00} \equiv - t_j$ and $Y_j^{11} = Y_j^{01} = Y_j^{10} = 0$. To do so, we use an integral representation involving the Gamma function:
\begin{align}
    \xi^q = \frac{1}{\Gamma(-q)} \int\limits_{0}^\infty \dd t \; t^{-q-1} \, \me^{-t \xi} ,
\end{align}
which converges for $\mathrm{Re}\, q < 0$ assuming that $\xi > 0$. Let $n = 1$ for now. Then, making the identification $\xi \equiv M_{BC} (x)$ our master formula \eqref{eq:master} gives
\begin{align}\label{eq:173-mz}
    \left< M_{BC}(x)^q \right>_{\cal V} &= \frac{1}{\Gamma(-q)} \int\limits_0^\infty \dd t \; t^{-q-1} \, \mathbb{E}\, \mathrm{Det}^{-1} \left(1 + {\textstyle{\frac{t}{2}}} K(x)\right) .
\end{align}
Recall that the determinant is over the $r$-dimensional space of scattering states.

In the case of a natural number $q \in \mathbb{N}$, one has the simplification that one can expand both sides of \eqref{eq:master} in powers of $t$ to obtain
\begin{align}\label{eq:N1_BC}
  \left< M_{BC}(x)^q \right>_{\cal V} =
  2^{-q}\, \Gamma(1+q) \, \mathbb{E} \, h_q \big( K(x) \big) ,
\end{align}
where $h_q$ is the complete homogeneous symmetric polynomial of degree $q$; cf.\ Sect.\ \ref{sect:obs}. This formula is consistent with \eqref{eq:173-mz} by analytic continuation in $q$. As a matter of fact, it extends to complex $q$ if $h_q \equiv s_\lambda$ is understood in the sense of Eq.\ \eqref{eq:schur} with $\lambda_1 = q$ and $\lambda_2 = \ldots = \lambda_r = 0$.

The situation with a general number $n \ge 1$ of observation regions can be handled in the same way. Starting from our equality of generating functions,
\begin{align}
    \big\langle \me^{- \sum_{j=1}^n t_j M_{BC}(x_j) } \big\rangle_\mathcal{V} = \mathbb{E} \, \mathrm{Det}^{-1} \left( 1 + {\textstyle{\frac{1}{2}}} \sum\nolimits_j t_j K(x_j) \right) ,
\end{align}
one can produce a formula which expresses the correlator \eqref{eq:171-mz} in terms of pure and mixed traces of powers of the local wavefunction data $K(x_1), \ldots, K(x_n)$. Here we discuss only the case of a single point contact, say at the link $c$. Then $K$ has rank one, and after performing the integrals we get:
\begin{align}
  \label{eq:Npt_BC_1pc}
  \begin{split}
    f_{q_1,\dots,\,q_n}(c\,; x_1, \ldots, x_n) &=
    2^{-(q_1 + \ldots + q_n)} \Gamma(1 + q_1 + \ldots + q_n)\\
    &\times \mathbb{E}\, \big( K(x_1)^{q_1} \cdots K(x_n)^{q_n} \big)
  \end{split}
\end{align}
with $K(x_j) = \sum_{\ell \in \mathcal{R}_j} w_j(\ell) \vert \psi_c(\ell) \vert^2$. Thus in the case of a single point contact, the relation between vertex-model and network-model observables is particularly easy to state: one passes from one side of the correspondence to the other by replacing the bosonic highest-weight operator $M_{BC}(x)$ with the square $K(x)$ of the scattering state's absolute value and, at the same time, replacing the vertex-model statistical average (with an operator insertion for the point contact) with the disorder average.

\subsubsection{Fermionic observables}

We now turn briefly to the fermionic counterparts of the bosonic observables of the preceding section. For a set $p_1, \ldots, p_n$ of non-negative integers, consider
\begin{align}\label{eq:g}
  g_{p_1,\, \ldots,\, p_n}(c\,; x_1, \ldots, x_n) &=
  \left< M_{FG}(x_1)^{p_1} \cdots M_{FG}(x_n)^{p_n}
  \right>_{\cal V} \\
  &= \STr_{\cal V} \pi(c) \rho(\widehat{U}_\mathrm{s})\,
  M_{FG}(x_1)^{p_1} \cdots M_{FG}(x_n)^{p_n} ,
\end{align}
with $M_{FG}(x_j) = \sum_{\ell \in \mathcal{R}_j} w_j(\ell)\, F_+(\ell) G_-(\ell)$. By setting $Y_j^{00} = Y_j^{01} = Y_j^{10} = 0$ and $Y_j^{11} = t_j$ one gets from Eq.\ \eqref{eq:master} the general formula
\begin{align}
    g_{p_1,\, \dots,\, p_n}(c\,; x_1,\dots,x_n) = \prod_{j=1}^n \frac{\partial^{p_j}}{\partial t_j^{p_j}} \bigg\vert_{t_j = 0}
    \mathbb{E} \Det\left(1 - {\textstyle{\frac{1}{2}}} \sum\nolimits_j t_j K(x_j) \right) .
\end{align}
Since the determinant is a polynomial in the parameters $t_j$ of degree $r$, there is a constraint (absent in the bosonic case) on the powers $p_j$ in order for the result to be non-trivial:
\begin{align}
    p_1 + \ldots + p_n \le r .
\end{align}
Thus, a non-trivial fermionic correlator with $n$ observation regions requires at least $n$ point contacts. In the special case of a single observation region ($n = 1$) one again obtains a very neat expression, now by the  elementary symmetric polynomial:
\begin{align} \label{eq:N1_FG}
  \left< M_{FG}(x)^p \right>_{\cal V} = \frac{p!}{(-2)^p} \, \mathbb{E}\, e_p \big( K(x) \big) .
\end{align}

\subsubsection{Weyl-symmetry relation}\label{sect:Weyl}

Here we specialize to the particular situation of a single observation region with just one link, $\ell$, in which case our matrix $K_{ik} = \bar\psi_{c_i} (\ell) \psi_{c_k}(\ell)$ has rank one with non-zero eigenvalue
\begin{align}
    \mathrm{Tr} \, K = \sum_{i=1}^r \vert \psi_{c_i} (\ell) \vert^2 \equiv A .
\end{align}
In this situation it is known \cite{Gruzberg2011} that there exists a remarkable relation,
\begin{align}\label{eq:Weyl}
    \mathbb{E} (A^q) = \mathbb{E} (A^{1-q}) ,
\end{align}
which has its foundation in a Weyl-group symmetry and holds exactly for any (finite) lattice and for any value of the scattering parameters $t_n$ of the time-evolution operator of the network model; cf.\ Eqs.\ (\ref{eq:5-mz}, \ref{eq:6-mz}). In fact, in order for the relation \eqref{eq:Weyl} to hold one only needs the property that the random phases of the network model are independent of each other and uniformly distributed \cite{Fyodorov2004}.

Our proof proceeds by showing that $\mathbb{E}(A^q)$ can be written as an integral
\begin{align}\label{eq:185-mz}
    \mathbb{E} (A^q) = \int_\mathbb{R} \mathrm{e}^{q \varphi} \omega(\varphi)\, \dd \varphi ,
\end{align}
where the function $\omega(\varphi)$ obeys the symmetry relation
\begin{align}\label{eq:186-mz}
    \mathrm{e}^{\varphi / 2} \omega(\varphi) =
    \mathrm{e}^{- \varphi / 2} \omega(- \varphi) .
\end{align}
The result \eqref{eq:Weyl} then follows immediately.

Thus, we need to establish Eqs.\ (\ref{eq:185-mz}, \ref{eq:186-mz}). We begin the calculation by applying the identity \eqref{eq:N1_BC} to translate the disorder average into a vertex-model statistical average:
\begin{align}
    \mathbb{E} (A^q) = \frac{2^q}{\Gamma(1+q)} \STr_{\cal V} \pi(c) \rho(\widehat{U}_\mathrm{s})\, \big( B_+(\ell) C_-(\ell) \big)^q .
\end{align}
Now, for the sake of the argument, imagine that we have carried out the supertrace $\STr_{\cal V}$ over all vertex-model states but those at the link $\ell$. What then remains is a final trace over the representation space $V \equiv V_\ell$ at the link $\ell$; cf.\ \eqref{eq:116-mz}:
\begin{align}\label{eq:onetrace}
    \mathbb{E} (A^q) = \frac{2^q}{\Gamma(1+q)} \,
    \mathrm{STr}_V \, \Omega \, (B_+ C_-)^q ,
\end{align}
where $B_+ = B_+(\ell)$, $C_- = C_-(\ell)$ and the operator $\Omega$ is the result of tracing the product $\pi(c) \rho(\widehat{U}_\mathrm{s})$ over all the non-$\ell$ degrees of freedom.

Next we convert the trace \eqref{eq:onetrace} into an integral over the (super-)coset space $G / K$ where $G = \mathrm{U}(1,1|2)$ is the (real) symmetry group of the SUSY vertex-model operator $\rho( \widehat{U}_\mathrm{s})$ and $K = \mathrm{U}(1|1) \times \mathrm{U}(1|1) \subset G$ the subgroup of symmetries of the Fock vacuum $\vert 0 \rangle $ and the point-contact operator $\pi(c)$. The conversion is done by a general formula
\begin{align}\label{eq:188-mz}
    \STr_V (O) = \int_{G/K} \dd g_K \left\langle 0 \vert \rho(g)^{-1} {O}\, \rho(g) \vert 0 \right\rangle ,
\end{align}
which is known from the theory of spin-coherent states (here: highest-weight vectors of the $G$-representation space $V$ built on the $K$-invariant Fock vacuum). The symbol $\dd g_K$ stands for the invariant Berezin integration form on $G/K$, and $g \mapsto \rho(g)$ is the representation \eqref{eq:rho} restricted to the link $\ell$.

In view of \eqref{eq:onetrace} we need to compute $\rho(g)^{-1} (B_+ C_-)^q \rho(g)$ for a well-chosen representative $g$ of the coset $gK \in G/K$. For this we parametrize $\rho(g)$ as $\rho(g) = \rho(n) \, \mathrm{e}^{\varphi H / 2}$ with $H$ the operator in \eqref{eq:H-mz} and $\rho(n)$ in the centralizer of $B_+ C_-$ (such a parametrization does exist). From the commutation relation \eqref{eq:sl2-mz} we then obtain
\begin{align}
    \rho(g)^{-1} (B_+ C_-)^q \rho(g) = \mathrm{e}^{- q \varphi} (B_+ C_-)^q .
\end{align}
Altogether, we now have for $\mathbb{E}(A^q)$ the integral representation
\begin{align}\label{eq:int-rep}
    \mathbb{E}(A^q) = \int_{G/K} \dd g_K \, \me^{-q \varphi(g)} \Phi(g) ,
\end{align}
where $\Phi(g)$ is a modified coherent-state expectation value:
\begin{align}
    \Phi(g) &= \frac{2^q}{\Gamma(1+q)} \left\langle 0 \vert \rho(g)^{-1} \Omega\, \rho(g) (B_+ C_-)^q \vert 0 \right\rangle \\
    &= \frac{1}{\Gamma(1+q) \Gamma(-q)} \int_0^\infty \dd t \, t^{-q-1} \left\langle 0 \vert \rho(g)^{-1} \Omega\, \rho(g)\, \me^{-2 t B_+ C_-} \vert 0 \right\rangle . \label{eq:193-mz}
\end{align}
Next, the operator $\me^{-2 t B_+ C_-}$ in the integral formula for $\Phi(g)$ has the following Gauss decomposition w.r.t.\ the original (particle-number) Cartan subalgebra:
\begin{align}
    \me^{-2t B_+ C_-} = \me^{ t(1+t)^{-1} \mathrm{e}^{-\mathrm{i}\alpha} b_+^\dagger b_-^\dagger}\, \me^{ - \log(1+t) (b_+^\dagger b_+^{\vphantom{\dagger}} + b_-^{\vphantom{\dagger}} b_-^\dagger)} \, \me^{ t(1+t)^{-1} \mathrm{e}^{\mathrm{i}\alpha} b_- b_+ }.
\end{align}
This identity is readily verified by multiplying the corresponding matrices in $\mathrm{SL}_2(\mathbb{C})$. By applying the Gauss decomposition to the Fock vacuum, we get
\begin{align}
    \me^{-2t B_+ C_-} \vert 0 \rangle = \me^{ t(1+t)^{-1} \mathrm{e}^{-\mathrm{i}\alpha} b_+^\dagger b_-^\dagger} \vert 0 \rangle \, (1+t)^{-1} .
\end{align}
When this is inserted into Eq.\ \eqref{eq:193-mz} the exponential disappears, since $\Phi(g)$ is independent of $\mathrm{e}^{\mathrm{i} \alpha}$ [recall Eq.\ \eqref{eq:int-rep}] and we may average over the parameter $\alpha$. By standard identities for the Gamma function, the $t$-integral now simply gives unity:
\begin{align}
    \frac{1}{\Gamma(1+q) \Gamma(-q)} \int_0^\infty \dd t \; t^{-q-1} (1+t)^{-1} = 1.
\end{align}
In this way we arrive at
\begin{align}
    \mathbb{E}(A^q) = \int_{G/K} \dd g_K \, \me^{-q \varphi(g)} \phi(g) , \quad \phi(g) = \langle 0 \vert \rho(g)^{-1} \Omega\, \rho(g) \vert 0 \rangle .
\end{align}
The key feature of the function $\phi(g)$ is its bi-invariance $\phi(k_1 g k_2) = \phi(g)$ for $k_1, k_2 \in K$, with the right invariance being part of the spin-coherent state construction and the left one stemming from the $K$-invariance of the operator $\pi(c) \rho(\widehat{U}_\mathrm{s})$.

Finally, changing integration variables as $g K = n a K$ and the corresponding invariant measures as $\dd g_K = \dd n\, \mathrm{e}^\varphi d\varphi$ with $a$ defined by $\rho(a) = \mathrm{e}^{\varphi H / 2}$, we obtain
\begin{align}
    \mathbb{E} (A^q) = \int_\mathbb{R} \mathrm{e}^{-q \varphi} \omega(-\varphi)\, \dd \varphi , \qquad \mathrm{e}^{-\varphi / 2} \omega(-\varphi) = \mathrm{e}^{\varphi / 2} \int \dd n\, \phi(n a),
\end{align}
which is Eq.\ \eqref{eq:185-mz} but for $\varphi \to -\varphi$. By the SUSY generalization \cite{Gruzberg2011} of a classical result due to Harish-Chandra, one has the Weyl-group symmetry \cite{Helgason}
\begin{align}
    \mathrm{e}^{\varphi/2} \int \dd n\, \phi(n a) =
    \mathrm{e}^{- \varphi/2} \int \dd n\, \phi(n a^{-1}) ,
\end{align}
which is the same as Eq.\ \eqref{eq:186-mz} and therefore completes our proof of \eqref{eq:Weyl}.


\section{Effective description by a Gaussian free field}
\label{sec:6}

In this section we will consider the continuum limit of the critical network model. Since the model under investigation does not fall into the category of quantum-integrable lattice models \cite{Zirnbauer1997}, there is little hope to obtain exact formulas on the lattice and take their limits. Nevertheless, building on the generally accepted hypothesis that the continuum limit is a conformal field theory (CFT), one may impose certain a priori constraints on the correlation functions. This, together with extensive numerical simulations, allowed us in \cite{BWZ} to derive several predictions concerning the continuum limit of the correlation functions discussed in Sect.\ \ref{sect:5.3}.

In the next subsection, we will review the results of \cite{BWZ} and point out that the Abelian operator product expansion (OPE) invoked there implies the description by a Gaussian free field, corresponding to a parabolic multifractality spectrum. This description is in agreement with our numerical studies, which confirm the abelian OPE and find unnaturally small deviations from parabolicity; see also the previous works {\cite{Evers2008b, Obuse2008}}. It should be mentioned that parabolicity of the spectrum was also proposed in a recent preprint \cite{Suslov2014} without any reference to CFT. Our reasoning leads to the same conclusion by relying on the well-established structure of conformal field theories.

\subsection{Review of previous results}\label{sect:6.1}

Our discussion here will focus on correlators of the bosonic highest-weight operators $M_{BC}^q$ only. It is for this sector of the theory that the constraints are most powerful. We will address possible generalizations at the end of the discussion.

When a critical lattice model is taken to the thermodynamic limit, one expects that lattice operators can be expanded in terms of CFT fields, and correlation functions can be computed within the framework of (perturbed) CFT. The global $\gl_{2|2}$-symmetry of the vertex model will now be used in order to classify fields of the CFT and select the ones to appear in the lattice expansions.
(Assuming that coarse graining is understood, we frequently write $B_+ C_-$ for $M_{BC}$.)

We start the analysis with some representation theory. While we have already discussed the representation-theoretic properties of $M_{BC}\,$, we still need to review those of $\pi = \pivac$. In contrast to our highest-weight vectors, $\pivac$ does not transform according to a single irreducible representation; instead, the following decomposition was established in \cite{Janssen1999}:
\begin{align}
  \pivac(c) = \int_{\frac{1}{2} + \mathrm{i} \mathbb{R}^+}
  \dd\mu(q_0)\, \langle V,V^*|q_0\rangle \, \phi_{q_0}^{\text{lat}}(c) .
\end{align}
The operators $\phi^{\text{lat}}_{q_0}$ are $K$-invariant elements in a principal series of $\gl_{2|2}$-irreducible infinite-dimensional representations labeled by $q_0\in \frac{1}{2}+i\mathbb{R}^+$, and $\dd \mu(q_0)$ and $\langle V,V^*|q_0\rangle$ are a known Plancherel measure and Clebsch-Gordan coefficients, which we do not need explicitly here. The relation between $\phi_q^{\text{lat}}$ and the highest-weight vector $(B_+C_-)^q$ is $\phi_q^{\text{lat}}$ can be obtained as the $K$-average of $(B_+C_-)^q$:
\begin{align}\label{eq:201-mz}
  \phi_q^{\text{lat}} = \int_{K} \dd k\, \rho(k) (B_+C_-)^q \rho(k)^{-1} ,
\end{align}
with $\dd k$ the Haar-Berezin measure on $K = \mathrm{U}(1|1) \times \mathrm{U} (1|1)$ and $\rho(k)$ the representation \eqref{eq:rho} restricted to a single link and to elements $k \in K$. (Note, however, that due to the presence of Grassmann derivations in the measure, the right-hand side is trivially zero when $q = 0$ or $q = 1$.)

Generic operators in the lattice model will have a similar decomposition over a continuum of representations. Let us then fully appreciate the fruit of all our labors: the highest-weight operators $M_{BC}^q\,$, transforming purely under the symmetry, will \underline{not} involve any integral over representations! Furthermore, they satisfy an Abelian fusion algebra of $\gl_{2|2}$-representations, and the combination of these aspects simplifies our discussion both on the lattice and in the continuum.

Before presenting the CFT predictions we stress that, trying to proceed constructively, we do not assume a priori the existence of a Kac--Moody symmetry in the CFT. Instead, we try to analyze the problem assuming no more than the well-accepted existence of a Virasoro algebra. In fact, it is known \cite{Read2001} that a subsector of the CFT for a localization problem in class $C$ does not possess a Kac-Moody symmetry, at least not the one extending the full global symmetry algebra. It is also known that the study of supersymmetric CFTs is complicated by non-unitarity and the presence of logarithmic fields \cite{Schomerus2006}. Note, however, that logarithms are expected to appear for atypical representations and not to appear for the typical (or generic) representations which underlie our correlation functions. (See \cite{Romain} for a recent discussion of logarithmic operators in localization problems.)

Let us now denote by $V_q$ the leading field in the expansion of $(B_+ C_-)^q$ in terms of CFT fields:
\begin{align}\label{eq:BCV}
  (B_+ C_-)^q(\ell) = \epsilon^{\Delta_{q}}
  V_{q}(z,\bar{z}) + o( \epsilon^{\Delta_{q}}) .
\end{align}
Here $\Delta_q$ is the scaling dimension of $V_q$ and $z = z(\ell)$ the complex coordinate of the link $\ell$ on the grid of mesh size $\epsilon$. We will sometimes call $V_q$ a vertex operator. A selection rule is that $V_q$ and the subleading terms have to transform in the same way as the operator on the left-hand side under the action of the $\gl_{2|2}$-symmetry. The lattice fields $\phi_q^{\text{lat}}$ will admit a similar expansion by CFT fields, and we denote by $\phi_q$ the leading term. By the relation \eqref{eq:201-mz}, $\phi_q$ has the same scaling dimension $\Delta_q\,$.

In \cite{BWZ} we put forth two conjectures: (i) $V_{q}$ and $\phi_q$ are spinless Virasoro primary fields, and (ii) the operators $V_q$ are closed under the fusion of Virasoro representations: $V_{q_1} \times V_{q_2} = V_{q_1 + q_2}\,$. At present, the only reliable way for us to test these conjectures is numerics. Now the first one leads to the following prediction for the correlator on the infinite plane:
\begin{align}\label{eq:203-mz}
  \left< \pi(c)(B_+ C_-)^q (\ell)\right> = \epsilon^{2\Delta_{q}} f_q \, |z_0-z|^{-2\Delta_{q}} + o(\epsilon^{2\Delta_{q}}), \quad
  q \in {\textstyle{\frac{1}{2}}} + \mathrm{i} \mathbb{R} ,
\end{align}
for some constant $f_q$ and $z_0 = z_0(c)$ the coordinate of the point contact; and we adapted our notation for correlation functions,
\begin{align}
    \left< \pivac(c) (B_+C_-)^q(\ell)\right> \equiv
    \left< (B_+ C_-)^q(\ell)\right>_\mathcal{V} \,,
\end{align}
in order to bring the point-contact operator $\pivac$ out into the open. Note that because of the change of Cartan subalgebra made in the construction of the highest-weight vectors, no constraint on the ``charge'' $q$ emerges from $\gl_{2|2}$-symmetry. Instead, it is the emerging conformal symmetry that selects the single component with dimension $\Delta_q$ out of the continuum of operators $\phi_q$ entering the decomposition of $\pivac\,$. In fact, this follows already from the weaker assumption that $\phi_q$ and $V_q$ are quasi-primaries \cite{DiFrancesco1997}. Our assumption (i) that the fields are spinless primaries leads to a prediction for the infinite cylinder as well as any other geometry which can be obtained by a conformal map from the plane. It is the cylinder prediction that was extensively checked numerically in \cite{BWZ}.

Once the pure scaling behavior is assumed, the exact symmetry relation of Sect.\ \ref{sect:Weyl} implies the same symmetry relation for the scaling dimensions:
\begin{align}
    \Delta_q = \Delta_{1-q} \,.
\end{align}
In particular, $\Delta_1 = \Delta_0 = 0$. We also remark that while the formulas initially were derived for $q \in \frac{1}{2} + \mathrm{i} \mathbb{R}$ only, the results make sense for other complex/real values of $q$.

Similar to \eqref{eq:203-mz}, another consequence of conjecture (i) is that the subleading terms entering the lattice expansion \eqref{eq:BCV} are descendant fields obtained by acting on $V_q$ with the Virasoro algebra. Their scaling dimensions are $\Delta_q + n$ with $n$ a positive integer. We caution, however, that this is expected to fail for $q = 1$ because of the presence of the (highest-weight component of the) $\gl_{2|2}$ current and its descendants in the expansion of the lattice operator. (The same happens for the well-understood case of vertex models of compact type; see \cite{Bondesan2015} for a recent discussion.) Actually, we could replace $B_+ C_-$ by the more general highest-weight vector $M_{BC}$ of Sect.\ \ref{ref:nilpotent}, whose flexible definition by a weight function $w$ allows one to choose the weights so as to eliminate the current. We refrain from elaborating this issue here and henceforth consider the case of generic $q$.

Turning to the operator product expansion (OPE), a first remark is that by the $\gl_{2|2}$-highest weight property, $V_{q_1 + q_2}$ is the only field among the $V_q$ that can appear in the OPE of $V_{q_1} \times V_{q_2}\,$. Our second conjecture, (ii), says that no other primary field intervenes, at least not so when correlators with an insertion only of the point-contact operator are considered. Hence, we have the formula
\begin{align}
  \label{eq:abelianOPE}
  V_{q_1}(z_1,\overline{z}_1) V_{q_2}(z_2,\overline{z}_2) =
  |z_{12}|^{\Delta_{q_1+q_2} - \Delta_{q_1} - \Delta_{q_2}}
  V_{q_1 + q_2}(z_2,\overline{z}_2) + \ldots
\end{align}
with $z_{12} = z_1 - z_2$ and the subleading terms being Virasoro descendants only.

Using Eq.\ \eqref{eq:abelianOPE}, one can reduce any correlator involving a string of operators $\prod_{i=1}^n V_{q_i}$ to one involving only $V_{q_1 + \ldots + q_n}$ and its descendants. This shows that $\pivac$ contributes a single scaling dimension $\Delta_{q_0}\,$, with
\begin{align} \label{eq:q0}
  q_0 = 1 - (q_1 + \ldots + q_n) ,
\end{align}
in all multi-point functions of the form \eqref{eq:171-mz}. While an equivalent choice for $q_0$ in \eqref{eq:q0} would be $q_0 = \sum q_i\,$, the present definition of $q_0$ lets us think about the point contact as contributing a charge $q_0$ that makes the total sum of charges $q_0 + q_1 + \ldots + q_n$ equal to the background value of $Q = 1$.

For consistency, the OPE between a vertex operator $V_{q_i}$ and the point contact can only involve one fusion channel, this time that of dimension $\Delta_{q_0 + q_i}$ with $q_0$ as above. When specialized to a three-point function on the plane, the abelian OPE implies
\begin{align}
  \begin{split}
    &\langle \pi(c)(B_+C_-)^{q_1}(\ell_1)(B_+C_-)^{q_2}(\ell_2)\rangle
    = \epsilon^{\Delta_{q_0}+\Delta_{q_1}+\Delta_{q_2}}
    f_{q_1,q_2}  |z_0-z_1|^{\Delta_{q_2}-\Delta_{q_0}-\Delta_{q_1}}\\
    &\times
    |z_0-z_2|^{\Delta_{q_1}-\Delta_{q_0}-\Delta_{q_2}}
    |z_1-z_2|^{\Delta_{q_0}-\Delta_{q_1}-\Delta_{q_2}}
    + o(\epsilon^{\Delta_{q_0}+\Delta_{q_1}+\Delta_{q_2}}) ,
  \end{split}
\end{align}
where as before, $q_0 = 1-(q_1+q_2)$, and $\{z_0,z_{1},z_2\}$ are the coordinates of $\{c,\ell_{1},\ell_2\}$. This prediction has also been tested and confirmed numerically in \cite{BWZ}.

Let us also point out another direct consequence of the abelian OPE, which is that in spite of the relation $\Delta_q = \Delta_{1-q}\,$, there is no such relation for the vertex operator,
\begin{align}
\label{eq:Vq_V1-q}
    V_q(z,\bar{z}) \neq R(q)\, V_{1-q}(z,\bar{z}) ,
\end{align}
if $R(q)$ is a function of $q$ only. This is so even though the representations indexed by $q = 1/2 + \mathrm{i}\lambda$ and $1-q = 1/2 - \mathrm{i}\lambda$ for $\lambda \in \mathbb{R}$ are equivalent, i.e., there exists a $\gl_{2|2}$-equivariant isomorphism intertwining them. (The latter generalizes the known equivalence of the corresponding $\su_{1,1}$-representations, which can be found e.g.\ in \cite{Vilenkin}.) To avoid running into a contradiction, one must now appreciate the following loop hole: our highest-weight vertex operators $V_q$ [or their lattice parents $(B_+ C_-)^q$], unlike the $K$-invariant vectors $\phi_q$ (or $\phi_q^{\rm lat}$), are not vectors in the $L^2$-space of the $\gl_{2|2}$-representation; rather, they are \emph{distributions} in the Schwartz sense of continuous linear functions on vectors. More concretely, if $X$ is a compact linear operator (such as the vacuum projector $X = \pivac$) acting on the $\gl_{2|2}$-representation space $V$, then the continuous linear mapping $X \mapsto \STr_V X (B_+ C_-)^q$ for $q = 1/2 \pm \mathrm{i} \lambda$ picks out the component of $X \in \mathrm{End}(V) \simeq V \otimes V^\ast$ in a $\gl_{2|2}$-irreducible representation space $V_{|\lambda|} \subset V \otimes V^\ast$, although $(B_+ C_-)^q \in \mathrm{End}(V) \simeq V \otimes V^\ast$ itself is not a vector inside that space $V_{|\lambda|}$. The upshot is that in spite of the equality $\Delta_q = \Delta_{1-q}$ of scaling dimensions, the operators $(B_+ C_-)^q$ and $(B_+ C_-)^{1-q}$ are not proportional to each other, and this distinction carries over to the CFT-fields $V_q\,$.

From a Lagrangian perspective, what we are after is the conformal field theory of a single scalar field $\varphi$ such that $V_q(z,\bar{z}) = \mathrm{e}^{q \varphi(z,\bar{z})}$. (Such a field $\varphi$ exists due to $V_q \sim (B_+ C_-)^q$ and the positivity of $B_+ C_-\,$.) The Lagrangian should emerge from the full theory as an effective Lagrangian by integrating out all the other fields (in a procedure that replaces the point contact by an effective operator which is a function of $\varphi$ with properties dictated by conformal symmetry as discussed above). Even if we do not know the full theory and cannot do the exercise of integrating out the fields, we can still make the following point: the inequality \eqref{eq:Vq_V1-q} means that there is no need for screening charges \cite{DiFrancesco1997} in a Coulomb-gas description of the problem. Thus there exists no obstruction for the correlators of $V_q$ to be those of a free theory. In fact, in the next section we will argue the case for a Gaussian free field starting from the postulate of the Abelian OPE.

\subsection{New predictions: multi-point functions and parabolicity}

Consider the following four-point function on the Riemann sphere:
\begin{align}\label{eq:general}
  G(\{z_i,\bar{z}_i\}) =
  \left< \pi(z_0,\overline{z}_0)
    V_{q_1}(z_1,\overline{z}_1)
    V_{q_2}(z_2,\overline{z}_2)
    V_{q_3}(z_2,\overline{z}_3) \right> ,
\end{align}
which, according to the chain of arguments in this paper, controls the leading behavior of the wavefunction correlator
\begin{align}
  \mathbb{E}\, \left( |\psi_c(z_1)|^{2q_1}
  |\psi_c(z_2)|^{2q_2} |\psi_c(z_3)|^{2q_3} \right)
\end{align}
in the presence of a point contact $c$ at $z_0\,$. The aim here is to show that demanding crossing symmetry of this correlation function entails the scenario of a Gaussian free field, and in particular, a parabolic spectrum of multifractal dimensions $\Delta_q\,$. The derivation employs nothing but standard CFT reasoning and will closely follow \cite{Lewellen1989}.

For our purposes it is sufficient to consider the function
\begin{align}
  Y(x,\overline{x}) =
  |x|^{\Delta_{q_2}+\Delta_{q_3} -2\gamma/3}
  |1-x|^{\Delta_{q_1}+\Delta_{q_2} -2\gamma/3}
  \left< \pi\left| V_{q_1}(1,1) V_{q_2}(x,\overline{x}) \right| V_{q_3}
  \right> ,
\end{align}
where we set the complex stereographic coordinates $(z_0,z_1,z_3)$ to fixed values $(\infty, 1, 0)$, and we multiplied by a prefactor for later convenience. With $q_0 = 1 - (q_1 + q_2 + q_3)$ the exponent $\gamma$ is given by
\begin{align}
  \gamma = \sum_{i=0}^3 h_{q_i} \,,
\end{align}
where $h_q = \Delta_q / 2$ denotes the holomorphic dimension. Using the Abelian OPE and the holomorphic factorization into a chiral and anti-chiral Virasoro algebra \cite{DiFrancesco1997}, one has $Y(x,\overline{x}) = |F(x)|^2$. Moreover, $F(x)$ has singularities only at $x = 0, 1, \infty$. By the OPE given in \eqref{eq:abelianOPE} the leading behaviors are
\begin{align}
  F(x) &= x^{h_{q_2+q_3}-\gamma/3}[a_0+o(x)] \\
  &= (1-x)^{h_{q_1+q_2}-\gamma/3}[b_0+o(1-x)] \\
  &= (1/x)^{h_{q_1+q_3}-\gamma/3}[c_0+o(1/x)] ,
\end{align}
for some numbers $a_0, b_0, c_0\,$. This constrains $F(x)$ to be of the form
\begin{align}
  F(x) = x^{h_{q_2+q_3}-\gamma/3}
  (1-x)^{h_{q_1+q_2}-\gamma/3}\tilde{F}(x) ,
\end{align}
where $\tilde{F}$ is analytic in a neighborhood of $x = 0$ and $x = 1$, and for $x \to \infty$ behaves as $\tilde{F}(x) = x^M (c + o (1/x))$ with exponent
\begin{align}
    M = \gamma - h_{q_1 + q_2} - h_{q_2 + q_3} - h_{q_1 + q_3}\,.
\end{align}
Now $\tilde{F}$ cannot have a branch point at $x = \infty$, or else there would be a branch cut from $\infty$ to, say $0$, in conflict with the requirement of regularity at the origin. $M$ is therefore an integer and $\tilde{F}$ an entire function with a pole at $\infty$, i.e., a polynomial of order $M$. Since $h_q$ is a continuous function of $q$, the integer $M$ cannot vary with $q$, and by taking the $q_i$ to zero, its value is inferred to be $M = 0$. It follows that $\tilde{F}$ is constant. The vanishing of $M$ now amounts to a constraint on the scaling dimensions:
\begin{align}\label{eq:suslov}
  \Delta_{q_1+q_2+q_3} + \Delta_{q_1} + \Delta_{q_2} + \Delta_{q_3} -
  \Delta_{q_1+q_2} - \Delta_{q_2+q_3}- \Delta_{q_1+q_3} = 0 ,
\end{align}
where $\Delta_{q_0} = \Delta_{q_1 + q_2 + q_3}$ was used. This condition appeared in early works of the CFT literature {\cite{Vafa1988, Lewellen1989}} and was first put forth in the Anderson-localization context in \cite{Suslov2014}. Given Eq.\ \eqref{eq:suslov}, the parabolic law for the multifractality spectrum $\Delta_q$ follows by setting $q_1 = q$ and $q_2 = q_3 = \eta$ and expanding in $\eta$, which at second order gives a constant second derivative $\Delta_q'' = \Delta_0''$. Since $\Delta_0 = \Delta_1 = 0$, this fixes $\Delta_q$ to be
\begin{align}\label{eq:parabol}
  \Delta_q = X q (1-q) ,
\end{align}
leaving one free parameter, $X$.

With the conformal block $F(x)$ determined, global conformal invariance allows one to reconstruct the four-point function $G(\{z_i, \bar{z}_i\})$ for arbitrary positions \cite{DiFrancesco1997}:
\begin{align}
  G(\{z_i,\bar{z}_i\}) \propto
  \prod_{i=1}^3 |z_{0i}|^{-2X (1 - q_1 - q_2 - q_3)\, q_i}
  \prod_{1 \le i<j \le 3} |z_{ij}|^{-2 X q_i q_j} .
\end{align}
Up to a prefactor, this coincides with the following correlator
\begin{align}
  \left< \delta(\varphi(z_0,\overline{z}_0))\,
    \mathrm{e}^{q_1 \varphi(z_1,\overline{z}_1)}
    \mathrm{e}^{q_2 \varphi(z_2,\overline{z}_2)}
    \mathrm{e}^{q_3 \varphi(z_3,\overline{z}_3)}
  \right>_{\text{GFF}} ,
\end{align}
where the expectation value is taken in a Gaussian free field (GFF) theory with action
\begin{align} \label{eq:S_phi}
  S = \frac{1}{8\pi X} \int \dd^2 z \sqrt{|g|}\,
  \big( (\partial_\mu \varphi)(\partial^\mu \varphi)
  + Q X \varphi R \big) .
\end{align}
Here $R$ is the curvature of the metric $g$, the number $Q$ is the background charge (in our case: $Q = 1$), and $X$ is the stiffness constant of the boson which determines the prefactor in the parabolic law for the scaling dimensions. The results of numerical studies \cite{BWZ} are compatible with the value $X = 1/4$.

We remark that the computation of the four-point function allows us to
fix the coefficients of the OPE of two fields $V_{q_1} V_{q_2}$ as that
of two vertex operators $\mathrm{e}^{q_1\varphi} \mathrm{e}^{q_2\varphi}$, with $\varphi$ the free field above. Thus we can identify generic multi-point
correlators with those of a free boson and produce a prediction for
correlators on the plane:
\begin{align}
  \mathbb{E}\, \big( |\psi_c(z_1)|^{2q_1} \cdots |\psi_c(z_n)|^{2q_n} \big)
  &\sim \left< \delta(\varphi(z_0,\overline{z}_0))\,
    \mathrm{e}^{q_1\varphi(z_1,\bar{z}_1)} \cdots
    \mathrm{e}^{q_n\varphi(z_n,\bar{z}_n)} \right>_\mathrm{GFF} \\
  &\propto \prod_{j}|z_0 - z_{j}|^{-2Xq_j(1-\sum_i q_i )}
  \prod_{i<j}|z_{i}-z_j|^{-2Xq_i q_j}\, ,
\end{align}
where $\sim$ has to be understood as the leading behavior in the continuum limit.

Let us mention that the free-boson effective theory \eqref{eq:S_phi} has central charge $c = 1 + 3 Q^2 X$. As indicated at the end of Sect.\ \ref{sect:6.1} this should be interpreted as the result of integrating out all other fields in an unknown full CFT, and it is not in contradiction with the fact that the partition function of the full theory is unity.

Let us add some perspective to our result. In a unitary CFT (or unitary QFT, for that matter) one has a Hermitian scalar product and thus the structure of a Hilbert space, with the vacuum and the excited states all being vectors in that Hilbert space. It should be clear by now that this standard mathematical framework is not appropriate for the SUSY vertex model at hand. Rather, the proper framework is the more general setting of statistical mechanics where one has a space of observables and a dual space of states or Gibbs measures, and statistical expectation values are given by the pairing between observables and states. To put it concretely, our observables $V_q \sim \mathrm{e}^{q \varphi}$ are unbounded (!) operators, and in order to get sensible correlation functions one must take the trace of these unbounded observables against states or density matrices that are regularized by the insertion of a compact operator such as $\pivac\,$.

One may be tempted to compare our GFF scenario with that of the $H_+^3$ Wess-Zumino-Witten model or of quantum Liouville theory, but when making such comparisons one should be aware of some important differences. On the one hand, (i) the $H_+^3$ WZW model and Liouville theory do not possess an invariant ground state, (ii) the one-point function $\langle V_q(x) \rangle$ does not exist, (iii) the two-point function $\langle V_{q_1} (x_1) V_{q_2}(x_2) \rangle$ exists only in the distributional sense, but (iv) all $n$-point functions for $n \geq 3$ exist. In contrast, (i) our SUSY vertex model does have a $\gl_{2|2}$-invariant ground state due to the presence of the fermionic degrees of freedom, (ii) the one-point function exists (and is zero, i.e., spontaneous symmetry breaking does not occur), (ii) the two-point function still exists in the distributional sense, but (iv) the $n$-point functions of the vertex operators $V_q$ do not exist for any $n \geq 3$ (all these properties are easily checked in the quasi-1D limit). As we have seen, what exists are $n$-point functions of the $V_q$ with the insertion of a point contact. Alternatively, one may regularize the theory by the presence of a conducting boundary or an absorbing background. (Incidentally, we feel that the important issue of regularization and how this might or might not invalidate the argument leading to the GFF description, is not treated in convincing fashion in \cite{Suslov2014}.)


\section{Conclusion}\label{sect:7}

In this work and its companion \cite{BWZ} we have developed a systematic approach to the Chalker-Coddington network model of the integer quantum Hall plateau transition. Our approach is based on a symmetry analysis of the equivalent vertex model, whose foundations were spelled out in Sects.\ \ref{sect:2}-\ref{sect:5}. While some of the formulas derived here were known in the literature, the substantial material regarding highest-weight operators of the non-compact sector of the vertex model and their correlation functions is new. A key feature of these unbounded operators is that they give rise to primary fields of the CFT albeit not of $L^2$-type. Our formalism allowed us to make contact with CFT correlation functions and to obtain several new results. A major result of this work is the derivation of a free-boson effective field theory. This is based on the abelian OPE conjectured and verified numerically in \cite{BWZ}. In particular, it gives an affirmative answer to the long-standing question of whether the spectrum of multi-fractal dimensions at the IQHE transition (and similar Anderson transitions) is parabolic.

One may wonder why numerical studies of the multifractality spectrum have so much difficulty \cite{Evers2008b,Obuse2008} seeing convergence to the parabolic law \eqref{eq:parabol}. Most likely, the reason is the presence in the conformal fixed-point theory of an irrelevant perturbation which is close to marginal (or even marginal), thereby causing corrections that disappear very slowly (perhaps only logarithmically) with increasing system size. Hence, any future numerical study using finite-size scaling will have to exercise the greatest possible care in order to ascertain that the scaling limit has (or has not been) reached.

The focus in this paper was solely on the Chalker-Coddington network model. Nonetheless, it should be clear that our concepts are of more general applicability. The emerging picture is that some (if not all) Anderson transitions in two dimensions where our methods apply, contain a free-boson sector describing certain correlation functions of critical wave functions. In each such case, the multifractality spectrum $\Delta_q$ will correspond to that of vertex operators which are exponentials of the free boson field, and by the symmetry relation $\Delta_q = \Delta_{q_* - q}$ \cite{Gruzberg2011}, it will be determined up to the stiffness constant for the boson. We also stress that a similar description by free bosons can be derived not only for the observables of Sect.\ \ref{sec:6}, but for the more general class considered in \cite{BWZ}. This, together with the generalized symmetry relations of \cite{Gruzberg2013}, allows one to fix the scaling dimensions of all these operators up to a single number. To take this reasoning further, we are currently investigating the transcription of our work to the network model of class $C$ (also known as a spin quantum Hall system). This is a promising direction, since a compact-target subsector of that model has been solved exactly \cite{Gruzberg1999}.

From the physics standpoint, the free-boson correlators discussed in this paper describe wavefunction observables at the critical point of an Anderson transition. To predict other characteristics such as the critical exponent of the localization length, an understanding of the full theory is required. It thus remains highly desirable to derive the full theory, of which the free boson elucidated in this paper is just one subsector.

\section*{Acknowledgment}

We thank I.\ Gruzberg and H.\ Saleur for useful discussions. Over the course of this work R.B.\ was supported by the DFG grant ZI 513/2-1 (and so was D.W.) and also by the EPSRC grants EP/I031014/1 and EP/N01930X/1.


\end{document}